\newif\iftoomuchdetail
\newif\ifuseprd
\newif\ifeprint
\newif\ifdatelast

\eprinttrue
\useprdtrue
\datelastfalse
\toomuchdetailfalse

\newif\ifrr
\rrfalse

\ifuseprd 
\documentclass[\ifeprint preprint,\fi
superscriptaddress,
tightenlines,%
floats,prd,eqsecnum,nobibnotes%
,nofootinbib,aps,12pt]{revtex4}
\else
\documentclass[11pt,letterpaper]{JHEP3}
\fi 
\usepackage{amsmath,amssymb}
\usepackage[final]{epsfig}
\allowdisplaybreaks[3]

\newcounter{saveequation}
\newcounter{detailnum}
\newcommand\savetheequation{\theequation}
\newcommand\detailtheequation{%
          $\delta$\Roman{detailnum}:\roman{equation}}

\newenvironment{detail}{\iftoomuchdetail\sf
         \setcounter{saveequation}{\value{equation}}%
         \setcounter{equation}{0}\addtocounter{detailnum}{1}%
         \renewcommand\theequation\detailtheequation%
         \fi}{
     \iftoomuchdetail%
     \ifnum\value{equation}=0\addtocounter{detailnum}{-1}\fi%
     \setcounter{equation}{\value{saveequation}}%
     \renewcommand\theequation\savetheequation%
     \fi%
     }

\ifuseprd
\let\oldappendix\appendix
\renewcommand\appendix{\oldappendix%
    \renewcommand\theequation{\thesection.\arabic{equation}}%
    \renewcommand\theparagraph{\roman{paragraph}}}
\fi

\DeclareMathOperator{\im}{Im}
\DeclareMathOperator{\real}{Re}

\DeclareMathOperator{\sech}{sech}
\DeclareMathOperator{\csch}{csch}

\newcommand\p{\ensuremath{\partial}}
\newcommand\evalat[2]{\ensuremath{\left.{#1}\right|_{#2}}}
\newcommand\abs[1]{\ensuremath{\left\lvert{#1}\right\rvert}}
\newcommand\norm[1]{\ensuremath{\left\|{#1}\right\|}}

\newcommand\field[1]{{\ensuremath{\mathbb{{#1}}}}}
\newcommand\ZZ{{\field{Z}}}
\newcommand\ZC{{\field{C}}}
\newcommand\ZR{{\field{R}}}
\newcommand\order[1]{{\ensuremath{{\mathcal O}\left({#1}\right)}}}
\newcommand\vev[1]{{\ensuremath{\left\langle{#1}\right\rangle}}}
\newcommand\anti[2]{\ensuremath{\left\{{#1},{#2}\right\}}}
\newcommand\com[2]{\ensuremath{\left[{#1},{#2}\right]}}
\DeclareMathOperator{\Tr}{Tr}

\newcommand\mathone{{\rlap{\kern .25em l}1}}
\newcommand\one{{\ifmmode{\text{\mathone}}\else{\mathone}\fi}}

\newcommand\apr{{\ensuremath{{\alpha'}}}}

\newcommand\ket[1]{\ensuremath{\lvert{#1}\rangle}}
\newcommand\bra[1]{\ensuremath{\langle{#1}\rvert}}
\newcommand\braket[2]{\ensuremath{\langle{#1}\rvert{#2}\rangle}}

\providecommand\FIGURE[2][]{\begin{figure}[#1]\begin{center}{#2}\end{center}
                       \end{figure}}

\providecommand\putabstract[1]{\ifuseprd\begin{abstract}%
     \vspace*{\baselineskip} {#1}  \vspace*{\fill} \end{abstract}%
                           \else\abstract{{#1}}\fi}
\newcommand\citejournal[4]{{\ifuseprd\else\begingroup\em\fi {#4}%
     \ifuseprd\else\endgroup\fi {\bf {#1}}\ifdatelast, {#3} ({#2})\else%
     \ ({#2}) {#3}\fi}}

\providecommand\atmp[3]{{\citejournal{#1}{#2}{#3}{Adv.\ Theor.\ Math.\ Phys.\ }}}
\providecommand\plb[3]{{\citejournal{#1}{#2}{#3}{Phys.\ Lett.\ B }}}
\providecommand\npb[3]{{\citejournal{{\em B\/}#1}{#2}{#3}{Nucl.\ Phys.\ }}}
\providecommand\jhep[3]{{\citejournal{#1}{#2}{#3}{J.\ High Energy Phys.\ }}}
\providecommand\jcap[3]{{\citejournal{#1}{#2}{#3}{J.\ Cosmol.\ Astropart.\ Phys.\ }}}

\providecommand\ptp[3]{{\citejournal{#1}{#2}{#3}{Prog.\ Theor.\ Phys.\ }}}

\providecommand\mpla[3]{{\citejournal{{\em A\/}#1}{#2}{#3}{Mod.\ Phys.\ %
   Lett.\ }}}

\providecommand\ijmpa[3]{{\citejournal{A#1}{#2}{#3}{Int.\ J.\ Mod.\ Phys.\ }}}

\providecommand\citeprd[3]{{\citejournal{#1}{#2}{#3}{Phys.\ Rev.\ D }}}

\providecommand\prl[3]{{\citejournal{#1}{#2}{#3}{Phys.\ Rev.\ Lett.\ }}}

\providecommand\citepr[3]{{\citejournal{#1}{#2}{#3}{Phys.\ Rev.\ }}}
\providecommand\prep[3]{{\citejournal{#1}{#2}{#3}{Phys.\ Rept.\ }}}
\providecommand\arxiv[2]{{\ifuseprd{\eprint{{\ifeprint\tt\fi {#1}/{#2}}}}%
                \else{\tt {#1}/{#2}}\fi}}
\providecommand\parxiv[2]{{\ifuseprd\else\tt\fi [\arxiv{#1}{#2}]}}
\providecommand\hepth[1]{\arxiv{hep-th}{{#1}}}
\providecommand\hepph[1]{\arxiv{hep-ph}{{#1}}}
\providecommand\grqc[1]{\arxiv{gr-qc}{{#1}}}

\providecommand\pchaodyn[1]{\parxiv{chao-dyn}{{#1}}}

\newcommand\phepth[1]{{\ifuseprd\else\tt\fi [\hepth{#1}]}}
\newcommand\phepph[1]{{\ifuseprd\else\tt\fi [\hepph{#1}]}}
\newcommand\pgrqc[1]{{\ifuseprd\else\tt\fi [\grqc{#1}]}}

\newcommand\pastph[1]{\parxiv{astro-ph}{{#1}}}
\newcommand\ct[1]{{\ifeprint\ifuseprd{\em{#1}},\else{\sf {#1}},\fi\fi}}
\newcommand\bt[1]{{\em {#1}},}

\newcommand\skipthis[1]{{}}

\newcounter{subfig}[figure]
\newlength{\figlen}
\renewcommand\thesubfig{(\alph{subfig})}
\newcommand\subfig[2][]{\refstepcounter{subfig}%
     \settowidth{\figlen}{{#2}}
     \begin{tabular}{c} {#2}  \\ 
     {\footnotesize \begin{minipage}{\figlen}
      \begin{center} \thesubfig\\ {#1}%
      \end{center} \end{minipage}} \end{tabular}}

\newcommand\vk{{\ensuremath{\vec{k}}}}
\newcommand\half{\ensuremath{\frac{1}{2}}}
\newcommand\thalf{\ensuremath{\tfrac{1}{2}}}
\newcommand\nvo{\ensuremath{\ket{\tilde{0}}_1}}
\newcommand\nvt{\ensuremath{\ket{\tilde{0}}_2}}
\newcommand\TO{\ensuremath{{\mathcal T}}}
\newcommand\op{\ensuremath{{\mathcal O}}}

\newcommand\eg{{\em e.g.\/}}
\newcommand\ie{{\em i.e.\/}}
\newcommand\cf{{\em cf.\/}}

\ifuseprd
\ifeprint
\let\wasaffiliation\affiliation
\renewcommand\affiliation[1]{\wasaffiliation{\footnotesize #1}}
\fi
\begin{document} 
\fi 

\title{\ifuseprd\vspace*{\fill}\fi Alpha-Vacua, Black Holes, and AdS/CFT}
\ifuseprd
\author{Andrew Chamblin}
\affiliation{Somewhere in Bush Country,
specifically, \\
       Department of Physics \\
       102 Natural Sciences Bldg. \\
       University of Louisville \\
       Louisville, KY \ 40292 \\ USA\\\vspace*{\baselineskip}}
\author{Jeremy Michelson}\email{jeremy@pa,uky,edu}
\altaffiliation[Current address: ]{Department of Physics;
The Ohio State University;
1040 Physics Research Building;
191 West Woodruff Avenue;
Columbus, Ohio \ 43210-1117; USA
}
\affiliation{Department of Physics and Astronomy \\
       University of Kentucky \\
       600 Rose Street \\
       Lexington, KY \ 40506 \\ USA\\
       \vspace*{\fill}
}
\else 
\author{Andrew Chamblin
\\
Somewhere in Bush Country, specifically \\
       Department of Physics \\
       University of Louisville \\
       102 Natural Sciences Bldg. \\
       University of Louisville \\
       Louisville, KY \ 40292 \\ USA
}
\author{Jeremy Michelson\thanks{\tt jeremy@pa,uky,edu}
\fi 
\putabstract{The Schwarzschild, Schwarzschild-AdS, and Schwarzschild-de
Sitter solutions all admit freely acting discrete involutions which
commute with the continuous symmetries of the spacetimes.
Intuitively, these involutions correspond to the antipodal map of the
corresponding spacetimes.  In analogy with the ordinary de Sitter
example, this allows us to construct new vacua by performing a
Mottola-Allen transform on the modes associated with the
Hartle-Hawking, or Euclidean, vacuum.  These vacua are the
``alpha''-vacua for these black holes.
The causal structure of a typical black hole may ameliorate
certain difficulties which are encountered in the case of de Sitter
$\alpha$-vacua.
For Schwarzschild-AdS black holes, a Bogoliubov transformation which mixes
operators of the two boundary CFT's provides a construction
of the dual CFT $\alpha$-states.
Finally, we analyze the thermal properties of these vacua.
}

\preprint{\begin{minipage}{10em}UK/06-17\\
   {\tt hep-th/0610133}\end{minipage}}


\ifuseprd
\maketitle
\else
\begin{document}
\fi 

\ifuseprd
\ifeprint
\tableofcontents
\fi
\fi

\section{Introduction}
 
The $\alpha$-vacua~\cite{m,a} of a scalar field in
the de Sitter (dS) spacetime have generated considerable interest (for example,
Refs.~\cite{bms,kklss,el,el2,gl,gl2,bfh,chm,cm,ch,nh}).
At first glance, these vacua appear to be highly
unphysical because they exhibit rather sick behaviour in the UV.
However, the idea is that they still may be important for cosmology
because they may correspond to certain quantum gravity effects which
have yet to be fully understood. For example, it has been
suggested
that the value of
$\alpha$-may be imprinted in the cosmic microwave background.
Discussions include~\cite{cosmo1,cosmo2,daniel1,daniel2,kklss,bjm,cosmo3}
and other relevant papers are~\cite{bm1,bm2,bm3,bm4,ml}.  $\alpha$-Vacua
for spin~1/2 fields were discussed in~\cite{hcf}.

Although $\alpha$-vacua have an unorthodox short distance behaviour, 
they are natural in that they preserve all of the de Sitter isometries.
In 
this sense, there is no reason to prefer one value of $\alpha$ over any other
value.  That said,
the literature~\cite{el,el2,gl,gl2,kklss,bfh} contains
many reasons, besides short distance physics,
for preferring the Euclidean 
vacuum obtained by analytic continuation from Euclidean dS,
to the other $\alpha$-vacua.

It has been suggested that some of the objectionable characteristics
of $\alpha$-vacua can be avoided
by changing the time ordering prescription of operators, in an $\alpha$-dependent
way.\cite{ch,nh}  
However,\cite{gl2} the path integral naturally picks out the usual
time ordering prescription for the Feynman propagator.  It is therefore unclear
to us what a change in the time ordering would mean, or how it would be
natural.  Thus we stick with the usual meaning of time ordering.

In this work, we introduce the notion of $\alpha$-vacua for black hole
spacetimes.  We start by discussing Schwarzschild, but in fact the
construction works in fair generality.  
The black hole horizon allows us to evade many of the objections that have
been raised in the context of de Sitter $\alpha$-vacua.
(We restrict
attention to neutral, nonrotating black holes and do not address whether
$\alpha$-vacua can be defined for black holes whose
singularities are not spacelike.)  In particular, we show that
Schwarzschild-AdS solutions admit $\alpha$-vacua.

These vacua could have
profound implications for Schwarzschild-AdS and the AdS/CFT
correspondence.  This is because there are two distinct boundary
components, and therefore two distinct theories, which encode the bulk
geometry of Schwarzschild-AdS.  The $\alpha$-vacua explicitly
``correlate'' points in the two different theories.  Maldacena
\cite{juan} has suggested that it might be possible to realize the
entropy of Schwarzschild-AdS as an entanglement entropy between the
two theories.

We describe how the $\alpha$-vacua can be constructed not only on the AdS side
but also on the CFT side.  On the CFT side, one essentially has to perform
a Bogoliubov transformation that mixes operators of the two CFTs, while
continuing to trace over a single {\em boundary\/}.  
(This prescription necessarily differs from the prescription~\cite{bms}
used in dS/CFT.)
We show that this
modifies the CFT in accord with the AdS---and also introduces $\alpha$-dependence
into the entropy, at least for the na\"{\i}ve definition of the entropy.

In~\cite{bln,rt} $\alpha$-vacua were constructed for certain asymptotically
AdS black holes---not those considered in this paper---from the de
Sitter $\alpha$-vacua
by using the de Sitter slicing of AdS.  In particular,~\cite{rt} showed that 
most $\alpha$-vacua
defined in this way are rendered ill-defined by the black hole singularity;
of this family of vacua,
only the Euclidean vacuum is well-defined.  Our construction of
$\alpha$-vacua is different, and does not rely on a foliation of
the spacetime by de Sitter slices.

This paper is organized as follows.  In $\S$\ref{sec:inv} we discuss the
geometry of Schwarzschild, and find a freely acting
antipodal symmetry which allows us to define $\alpha$-vacua.  We extend this
to a much larger class of spacetimes, including Schwarzschild-AdS,
in $\S$\ref{sec:gen}.  $\alpha$-vacua are then defined and examined in
$\S$\ref{sec:alpha}.  Objections to $\alpha$-vacua for de Sitter are reviewed
in $\S$\ref{sec:object} and then rebutted for the spacetimes considered in
this paper.
Comments therein regarding
string theory amplitudes for $\alpha$-vacua in Rindler
space are elaborated on in Appendix~\ref{sec:rindler}.
The CFT-dual states to $\alpha$-vacua are described in $\S$\ref{sec:CFTalpha}.
Consequences of AdS/CFT are then described in $\S$\ref{sec:AdSCFT}.
We ensure that the CFT states, that we propose are dual to
AdS $\alpha$-vacua,
do indeed reproduce AdS propagators and particle production.
We close with the observation that black hole entropy is $\alpha$-dependent.

\subsection*{Note:} 
This project was initiated about two years ago with $\S$\ref{sec:inv},
which resulted from a conversation between Andrew, S.~Das and A.~Shapere
while Andrew was driving them in his Land Rover,
and Andrew was actively involved with all of the essential part of the work.
Unfortunately, Andrew passed
away while this work was waiting to be finalized.
J.M. thanks Andrew's friends, family and colleagues
for the opportunity to have
presented this work at the Andrew Chamblin Memorial Symposium in
March 2006 at the University of Louisville.  J.M. also apologizes
for the delay in producing this last paper of Andrew's---and thanks those
who encouraged J.M. to get Andrew's paper into final form.
Andrew will be missed.

\section{The Discrete Involutions of Schwarzschild} \label{sec:inv}
 
To begin, let $({\cal M}, g)$ denote the Schwarzschild spacetime. We
are interested in constructing discrete involutive isometries which
will act on ${\cal M}$.  In particular, we are interested in the
actions of time and space inversion. Since we wish to consider the
actions of these inversions on the maximally extended spacetime, it is
most natural to use Kruskal coordinates~\cite{kruskal}, which cover the entire
manifold.  We therefore begin with a review of the relation of these
coordinates to the usual Schwarzschild coordinates (which cover only
part of the maximal extension).
 
Let $(t, r, {\theta}, {\phi})$ denote the Schwarzschild coordinates
so that the metric reads
\begin{equation}
ds^{2} = -\left(1 - \frac{2m}{r}\right)dt^{2}
 \,+\, \frac{dr^{2}}{\left(1 - \frac{2m}{r}\right)}
 \,+\, r^{2}\left(d{\theta}^2
 \,+\, \sin^2 \theta \, d\phi^2\right) .
\end{equation}
Next introduce null coordinates $u$ and $v$ such that
\begin{align}
du &= dt - \frac{dr}{\left(1 - \frac{2m}{r}\right)}, &
dv &= dt + \frac{dr}{\left(1 - \frac{2m}{r}\right)},
\end{align}
or, integrating
\begin{align}
u &= t -r -2m \log (r - 2m) , &
v &= t + r + 2m \log (r - 2m).
\end{align}
Now form the coordinates $U$ and $V$ by exponentiating
\begin{align}
U &= - e^{-\frac{u}{4m}}, &
V &= e^{\frac{v}{4m}},
\end{align}
Then one finds that the coordinates $T$ and $Z$ defined by
\begin{align}
T &= \sinh \left(\frac{t}{4m}\right) e^{\frac{r}{4m}} \sqrt{r - 2m},
 &
Z &= \cosh \left(\frac{t}{4m}\right) e^{\frac{r}{4m}} \sqrt{r - 2m},
\end{align}
satisfy the simple algebraic relations
\begin{align}
U &= T \,-\, Z, &
V &= T \,+\, Z,
\end{align}
\ie, $U$ and $V$ are advanced and retarded null coordinates relative
to $T$ and $Z$. One checks that in these coordinates the metric
assumes the form
\begin{equation} \label{Schwds2}
ds^2 = e^{-\frac{r}{2m}}\left(16\frac{m^2}{r}\right)\left(-dT^2 \,+\,
dZ^2\right) \,+\,
r^2\left(d\theta^2 \,+\, \sin^2\theta\,d\phi^2\right) .
\end{equation}
Using the coordinates $(T, Z, \theta, \phi)$, we define total time
inversion by the map
\begin{equation} \label{RT}
R_{T}: \,(T, Z, \theta, \phi) \,\longrightarrow\, (-T, Z, \theta, \phi),
\end{equation}
and likewise space inversion is given by
\begin{equation} \label{RZ}
R_{Z}: \,(T, Z, \theta, \phi) \,\longrightarrow\, (T, -Z, \theta, \phi).
\end{equation}
Of course, neither of these involutions acts freely (they both have
fixed points). To obtain a free action, we need to take a product with
some other map which {\em is\/} freely acting. Such a map, which we
denote $P$, is given as
\begin{equation} \label{P}
P: \,(T, Z, \theta, \phi) \,\longrightarrow\, (T, Z, \pi - \theta, \phi + \pi).
\end{equation}

Thus, we can construct the following four freely acting involutions on
${\mathcal M}$: $P, PR_{T}, PR_{Z}$ and $PR_{Z}R_{T}$. We claim that
all of these involutions extend to the corresponding Euclidean
instanton (the ``cigar''). Before addressing the Riemannian issue,
however, we need to first determine which of these maps commutes with
the continuous symmetries of the spacetime.  This was actually studied
by Gibbons \cite{gazza}, who found that only the map 
$J = P R_Z R_T$ commutes with the continuous symmetries of 
${\mathcal M}$.
We shall refer to this antipodal map as the ``CPT'' operator.  If
we consider the {\em quotient} manifold
\begin{equation}
{\mathcal M}_J = {\mathcal M}/J
\end{equation}
then it is straightforward to show that ${\mathcal M}_J$ is
asymptotically flat, is {\em not\/} time orientable, and is space orientable.
In order to see how these involutions extend to the Euclidean section,
it is useful to write complexified Schwarzschild as an algebraic
variety in $\ZC^7$ in the usual fasion.
Explicitly, let $\{Z^i \,|\,i = 1,\dots,7\}$ be coordinates
on $\ZC^7$, so that in terms of Schwarzschild
coordinates (which cover only a subset of the variety), we have
\begin{equation}
\begin{aligned}
Z^1 &= r\sin \theta \cos \phi, \\
Z^2 &= r\sin \theta \sin \phi, \\
Z^3 &= r\cos \theta,
\end{aligned} \quad \begin{aligned}
Z^4 &= -2m \sqrt{\frac{2m}{r}} \,+\, 4m \sqrt{\frac{r}{2m}}, \\
Z^5 &= 2\sqrt{3} m \sqrt{\frac{2m}{r}},
\end{aligned} \quad \begin{aligned}
Z^6 &= 4m \sqrt{1 - \frac{2m}{r}} \cosh \left(\frac{t}{4m}\right), \\
Z^7 &= 4m \sqrt{1 - \frac{2m}{r}} \sinh \left(\frac{t}{4m}\right).
\end{aligned}
\end{equation}

With the coordinates as in~\eqref{Schwds2}, it turns out that complexified
Schwarzschild (${\mathcal M}_\ZC$) is given as the algebraic variety
determined by the three polynomials
\begin{equation}
\begin{aligned}
(Z^6)^2 - (Z^7)^2 \,+\, \frac{4}{3} (Z^5)^2 &= 16m^2, \\
\left[(Z^1)^2 \,+\, (Z^2)^2 \,+\, (Z^3)^2\right] (Z^5)^4 &= 576m^6 , \\
\sqrt{3} Z^4 Z^5 \,+\, (Z^5)^2 &= 24m^2.
\end{aligned}
\end{equation}

The Lorentzian section (${\mathcal M} = {\mathcal M}^L$)
and the Riemannian section (${\mathcal M}^{R}$)
are determined by finding certain anti-holomorphic involutions acting on
the above variety which stabilise either
${\mathcal M}^{L}$ or ${\mathcal M}^{R}$; that is, we find maps
\begin{align}
J_L: & \,{\mathcal M}_\ZC \,\longrightarrow\, {\mathcal M}_\ZC, &
J_R: & \,{\mathcal M}_\ZC \,\longrightarrow\, {\mathcal M}_\ZC,
\end{align}
such that $J_L$ leaves ${\mathcal M}^L \,\subset\, {\mathcal M}_\ZC$
invariant:
\begin{equation}
J_L ({\mathcal M}^L) = {\mathcal M}^L,
\end{equation}
and such that $J_R$ leaves ${\mathcal M}^R \,\subset\, {\mathcal M}_\ZC$
invariant:
\begin{equation}
J_R ({\mathcal M}^R) = {\mathcal M}^R.
\end{equation}

As described in \cite{cham}, $J_L$ restricted to ${\mathcal M}^L$ is an
anti-holomorphic version of time reversal.
$J_R$ is the map given by reflection through the
$\tau = 0$ (where $\tau = it$) three-surface in the ``cigar'' instanton 
(\ie, $\tau = 0$ is the ``Einstein Rosen bridge''
three-surface ${\Sigma}$, with topology
$S^2 \,\times\, \ZR$).
Since the surfaces $t = 0$ and $\tau = it = 0$
correspond to the surface $Z^7 = 0$, we see that
${\mathcal M}^L$ and ${\mathcal M}^R$
intersect precisely along this Einstein Rosen bridge. Explicitly, we
can realise the two maps $J_L$ and $J_R$ as follows:
\begin{equation}
\begin{aligned}
J_L: & \, (Z^1, Z^2, Z^3, Z^4, Z^5, Z^6, Z^7) \,\longrightarrow\,
(\bar{Z}^1, \bar{Z}^2, \bar{Z}^3, \bar{Z}^4, \bar{Z}^5, \bar{Z}^6, \bar{Z}^7),
\\
J_R: & \, (Z^1, Z^2, Z^3, Z^4, Z^5, Z^6, Z^7) \,\longrightarrow\,
(\bar{Z}^1, \bar{Z}^2, \bar{Z}^3, \bar{Z}^4, \bar{Z}^5, \bar{Z}^6, -\bar{Z}^7).
\end{aligned}
\end{equation}
Comparing these explicit formulae for $J_L$ and $J_R$ with the
coordinates in \eqref{Schwds2}, we see that $J_R$ is thus obtained from 
$J_L$ by the transformation $t \,\longrightarrow\, \tau = it$.

What we want to do now is show how the maps $R_{T}, R_{Z}$ and $P$
acting on ${\mathcal M}^L$, and likewise their Euclidean counterparts
acting on ${\mathcal M}^R$, are actually just the {\em restrictions}
to ${\mathcal M}^L$ and ${\mathcal M}^R$ of certain {\em holomorphic}
involutions acting on ${\mathcal M}_\ZC$. Of course, once
we notice that our complex coordinates $Z^6$ and $Z^7$ are (up to
a scaling) actually our Kruskal coordinates $Z$ and $T$, it is easy to
see that the ``big'' involutions, ${\mathcal R}_Z$ and ${\mathcal R}_T$
(which restrict to $R_Z$ and $R_T$ on ${\mathcal M}^L$)
are given by
\begin{align}
{\mathcal R}_Z: &\, (Z^1, Z^2, Z^3, Z^4, Z^5, Z^6, Z^7) \,\longrightarrow\,
   (Z^1, Z^2, Z^3, Z^4, Z^5, -Z^6, Z^7), \\
{\mathcal R}_T: &\, (Z^1, Z^2, Z^3, Z^4, Z^5, Z^6, Z^7) \,\longrightarrow\,
   (Z^1, Z^2, Z^3, Z^4, Z^5, Z^6, -Z^7).
\end{align}
Clearly, these maps are holomorphic,
and since they commute with both $J_L$ and $J_R$,
they restrict to well-defined involutions on ${\mathcal M}^L$ and
${\mathcal M}^R$. Thus, ${\mathcal R}_Z|_{{\mathcal M}^L} = R_Z$ and
${\mathcal R}_T|_{{\mathcal M}^L} = R_T$.
For the maps restricted to the Riemannian
section, we shall write
\begin{align}
{\mathcal R}_Z|_{{\mathcal M}^R} &= \bar{R}_Z: 
\, {\mathcal M}^R \,\longrightarrow\, {\mathcal M}^R \\
{\mathcal R}_T|_{{\mathcal M}^R} &= \bar{R}_T:
 \, {\mathcal M}^R \,\longrightarrow\, {\mathcal M}^R
\end{align}
In terms of local coordinates on ${\mathcal M}^R$,
these reflections take the form
\begin{align}
{\bar R_Z}: &\, \tau \,\longrightarrow\,  -\tau \,+\, 4\pi m, &
{\bar R_T}: &\, \tau \,\longrightarrow\, - \tau
\end{align}
($r, \theta$, and $\phi$ are left invariant by both these maps).
Thus, we see that $\bar{R}_T$ reflects in imaginary time whereas
$\bar{R}_Z$ corresponds to rotating through half a period in imaginary time.

Finally, we obtain the involution $\bar{P}$ on ${\mathcal M}^R$ by
restricting to ${\mathcal M}^R$ the following map on ${\mathcal M}^\ZC$:
\begin{equation}
{\mathcal P}: \, (Z^1, Z^2, Z^3, Z^4, Z^5, Z^6, Z^7) \,\longrightarrow\,
(-Z^1, -Z^2, -Z^3, Z^4, Z^5, Z^6, Z^7).
\end{equation}
                                                                                             
Now that we have made sense of how to extend our discrete isometries
$R_Z, R_T$ and $P$ from ${\mathcal M}$ to ${\mathcal M}^R$, we
can act on the Hartle-Hawking modes of the Kruskal manifold with these
involutions.  This allows us to apply the Mottola-Allen transformation
and so obtain $\alpha$-vacua.  It should be obvious that one may obtain
these Euclidean involutions for any spherically symmetric black hole
spacetime, and so we now turn our attention to the more general case.

\section{Geometry of a General, Neutral, Spherically Symmetric Black Hole}
\label{sec:gen}

Following~\cite{fhks}, we consider the $(d+1)$-dimensional
spacetime with metric
\begin{equation} \label{extds}
ds^2 = -f(r) dt^2 + \frac{dr^2}{f(r)} + r^2 d\Omega_{d-1}^2,
\end{equation}
where $f(r)$ is a monotonically increasing (for positive $r$) function with a
singularity at $r=0$ and a simple
zero at $r=r_+>0$.  The assumed monotonicity implies that $r=r_+$ is
the only real, positive zero.  The singularity
at $r=0$ is assumed to give a spacetime singularity, and so the
spacetime cannot be extended beyond $r=0$.  Examples are Schwarzschild
[$f(r)=1-\frac{\omega_d m}{r^{d-2}}$] and Schwarzschild-AdS
[$f(r)=\frac{r^2}{\ell^2}+1-\frac{\omega_d m}{r^{d-2}}$] where
$\ell$ is the length-scale associated with the
cosmological constant, and $m$ is the properly
normalized mass of the black hole, when $\omega_d=\frac{16\pi
G_N}{(d-1) \text{Vol}(S^{d-1})}$, with $G_N$ the Newton constant.

\iftoomuchdetail
\begin{detail}
The nonzero curvature objects are
\begin{gather}
\begin{aligned}
\Gamma^t_{rt} &= \frac{f'(r)}{2 f(r)}, &
\Gamma^r_{tt} &= \frac{1}{2} f(r) f'(r), &
\Gamma^r_{rr} &= -\frac{1}{2} \frac{f'(r)}{f(r)}, \\
\Gamma^\alpha_{r\beta} &= \frac{1}{r} \delta^\alpha_\beta, &
\Gamma^r_{\alpha\beta} &= -r f(r) \hat{g}_{\alpha\beta}, &
\Gamma^\alpha_{\beta\gamma} &= \hat{\Gamma}^\alpha_{\beta\gamma},
\end{aligned} \\
\begin{aligned}
R_{trtr} &= \frac{1}{2} f''(r), &
R_{t\alpha t\beta} &= \frac{1}{2} r f(r) f'(r) \hat{g}_{\alpha\beta}, &
R_{r\alpha r\beta} &= -\frac{1}{2} r \frac{f'(r)}{f(r)} \hat{g}_{\alpha\beta},
&
R_{\alpha\beta\gamma\delta} 
  &= 2 r^2 [1-f(r)] \hat{g}_{\alpha[\gamma} \hat{g}_{\delta]\beta},
\end{aligned} \\
\begin{gathered}
\begin{aligned}
R_{tt} &= \frac{1}{2} f(r) f''(r) + \frac{d-1}{2 r} f(r) f'(r), &
R_{rr} &= -\frac{f''(r)}{2 f(r)} - \frac{d-1}{2r} \frac{f'(r)}{f(r)},
\end{aligned} \\
R_{\alpha\beta} = (d-2)[1-f(r)] \hat{g}_{\alpha\beta}
 -r f'(r) \hat{g}_{\alpha\beta},
\end{gathered} \\
R = \frac{(d-1)(d-2)}{r^2}[1-f(r)] - \frac{2(d-1)}{r}f'(r) - f''(r),
\end{gather}
where $\alpha,\beta,\dots$ index coordinates on the sphere, and the
hat denotes the corresponding geometric quantities for the unit
sphere.  We have used
$\hat{R}_{\alpha\beta\gamma\delta}=2g_{\alpha[\gamma} g_{\delta]\beta}$.
\end{detail}%
\fi%

There is a coordinate singularity (horizon) at $f(r)=0$, \ie\
$r=r_+$.  We extend the coordinates past the horizon, and eventually
define Kruskal coordinates~\cite{kruskal} by first defining the
tortoise coordinates
$r^*$ by
\begin{equation}
r^*(r) = \int_0^r \frac{dr'}{f(r')} + \frac{\pi i}{f'(r_+)}.
\end{equation}
The rationale for the additional, imaginary constant is as follows.\cite{fhks}
We want the coordinate change to be well-defined in the region of
overlap with the original coordinates, namely $r>r_+$.  But the
integration from $r=0$ passes through the pole at $r=r_+$ and contributes
an imaginary part to the integral.  This has been explicitly cancelled
(with the contour going below the pole).
\iftoomuchdetail
\begin{detail}%
In particular, observe that
\begin{equation} \label{r*r+}
r^*(r\sim r_+) \sim \frac{1}{f'(r_+ )} \ln (r-r_+);
\end{equation}
this has an imaginary part for $r<r_+$.
\end{detail}%
\fi

Next define the double-null coordinates
\begin{align}
u &= t-r^*, &
v &= t+r^*.
\end{align}
The metric reads
\begin{equation}
ds^2 = -f(r) du dv + r^2 d\Omega_{d-1}^2,
\end{equation}
where, of course, $r$ is now considered a function of $u$ and $v$.
Since $f(r)$ was assumed to be monotonic, $r>0$ is, in fact, a single
valued function of $v-u$.  Clearly, however, this metric is still
singular at $r=r_+$.

With the conformal transformation
\begin{align}
U &= -e^{-\frac{f'(r_+)}{2} u} = T-Z, & V&=e^{\frac{f'(r_+)}{2} v} = T+Z,
\end{align}
the metric becomes
\begin{equation} \label{Kds}
\begin{split}
ds^2 &= -\frac{4}{f'(r_+)^2} f(r) e^{-f'(r_+) r^*} dU dV + r^2 d\Omega_{d-1}^2
\\   &= \frac{4}{f'(r_+)^2} f(r) e^{-f'(r_+) r^*} (-dT^2+dZ^2)
       + r^2 d\Omega_{d-1}^2,
\end{split}
\end{equation}
and we have the usual identities
\begin{align}
U V &= T^2-Z^2=- e^{f'(r_+) r^*}, & 
\tanh^{-1} \frac{T}{Z} &= \frac{f'(r_+)}{2} t.
\end{align}
\iftoomuchdetail
\begin{detail}%
Applying the approximation~\eqref{r*r+} for $r\sim r_+$, we see
that t%
\end{detail}%
\else
T%
\fi
he metric is now smooth at $r=r_+$, and therefore the $U,V$
coordinates extend to the singularity at $r=0$.  These are the Kruskal
coordinates%
\iftoomuchdetail
\else 
\ from which one can easily obtain the Penrose diagram%
\fi
.  The shape of the Penrose diagram depends on $r^*(r=\infty)$, 
as shown in Figure~\ref{fig:PD}.%
\iftoomuchdetail
\else 
\footnote{The authors of~\cite{fhks} use rescaled Penrose coordinates
  relative to us, and so obtain bowed singularities instead of bowed
  asymptotics.  However, this rescaling is not possible for asymptotically flat
  spacetimes, so we have consistently drawn bowed asymptotics.}
\fi

\FIGURE[t]{ 
\begin{tabular}{cc}
\subfig{\includegraphics[width=3in,clip=true
]{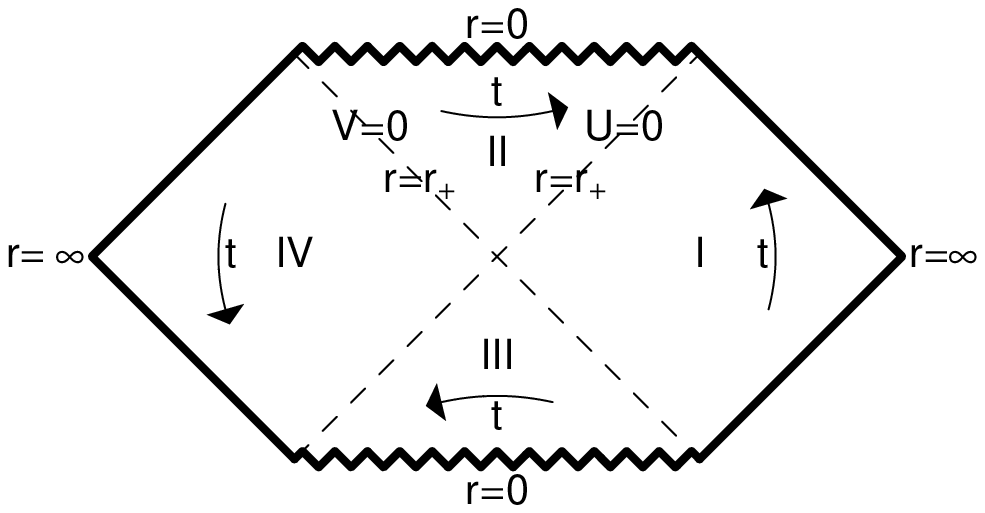}}
    \label{fig:PD:schwarz} &
\subfig{\includegraphics[width=3in,clip=true
]{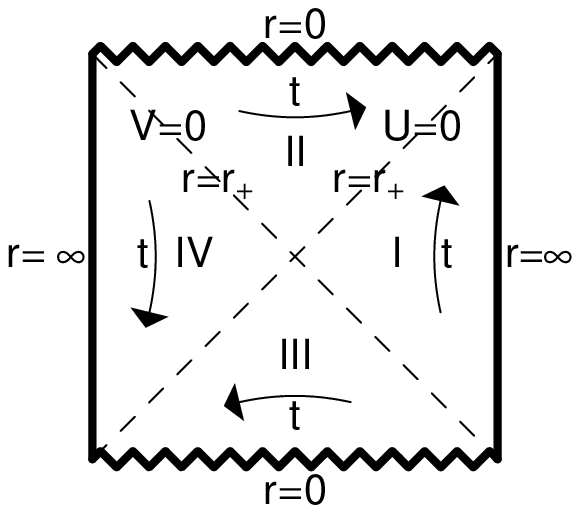}}
    \label{fig:PD:AdS} \\
\subfig{\includegraphics[width=3in,clip=true
]%
    {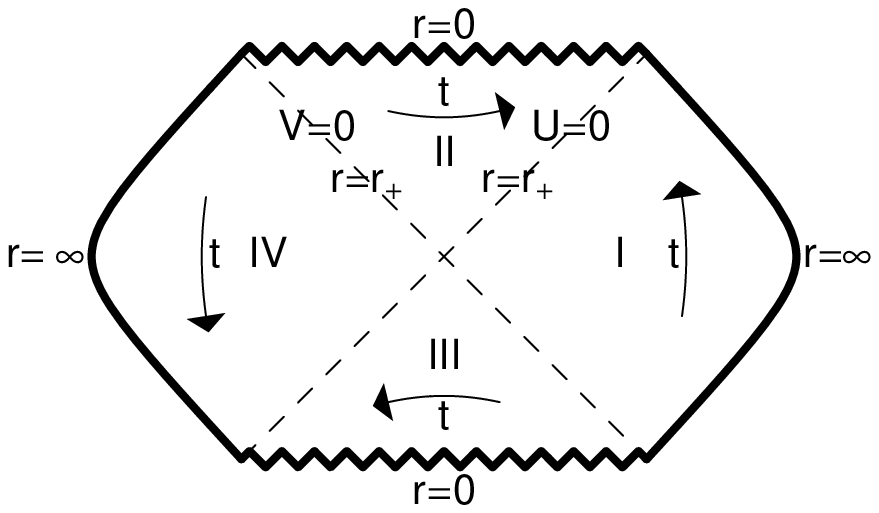}} \label{fig:PD:AdSs} &
\subfig{\includegraphics[width=3in,clip=true
]{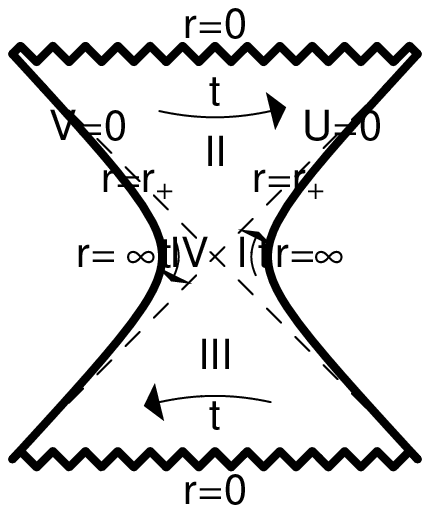}}
    \label{fig:PD:bowin}
\end{tabular}
\caption{Penrose diagrams of the various types of spacetimes.  The
exterior and interior regions have been labelled in the usual way.
The direction of time, Eq.~\eqref{time}, in each region is also indicated.
\ref{fig:PD:schwarz}~\hbox{$r^*(r=\infty)=\infty$} ({\em e.g.\/}
  asymptotically flat spacetimes.)
\ref{fig:PD:AdS}~\hbox{$r^*(r=\infty)=0$} (\eg\ the BTZ black hole).
\ref{fig:PD:AdSs}~\hbox{$0<r^*(r=\infty)<\infty$}.  Large mass
AdS$_5$-Schwarzschild is the example drawn,
for which \hbox{$r^*(r=\infty)=\pi$}.
\ref{fig:PD:bowin}~\hbox{$r^*(r=\infty)<0$}.  The
figure is for \hbox{$r^*(r=\infty)=-\pi$}.%
\label{fig:PD}}
}

\iftoomuchdetail
\begin{detail}%
To draw a Penrose diagram, define
\begin{align}
\tilde{U} &= \tan^{-1} U, & \tilde{V} &= \tan^{-1} V.
\end{align}
\iftoomuchdetail
\begin{detail}%
Clearly, $\tilde{U},\tilde{V}$ lie between $-\frac{\pi}{2}$ and
$\frac{\pi}{2}$, subject to the condition that $r>0$.  In particular, the
singularity at $r=0$ ($r^* = \frac{\pi i}{f'(r_+)}$) is located at 
$UV=1$, or $\tilde{U} = \pm \frac{\pi}{2} - \tilde{V}$.
The na\"{\i}ve
constraints on the range of $\tilde{U},\tilde{V}$ imply that the sign
correlates with $\tilde{V}\gtrless 0$.
On the Penrose
diagram, $\tilde{T}=\frac{1}{2}(\tilde{U}+\tilde{V})$,
$\tilde{Z}=\frac{1}{2} (\tilde{V}-\tilde{U})$, these are (parallel
horizontal) lines of
constant $\tilde{T}=\pm \frac{\pi}{4}$, with
$-\frac{\pi}{4}<\tilde{Z}<\frac{\pi}{4}$, just as for Schwarzschild.

\end{detail}%
\fi
The horizon is located at $r=r_+$, or, from the
behaviour~\eqref{r*r+}, $\real r^*=-\infty$, and thus $UV=0$.
\iftoomuchdetail
\begin{detail}%
These are
the null lines $\tilde{U}=0$ and $\tilde{V}=0$, again just as for
Schwarzschild.
Finally, t%
\end{detail}%
\else
T%
\fi
here are two possibilities for the boundary at $r=\infty$.

If $r^*(r=\infty)$ diverges---that is, if $f(r)$ asymptotes to a
constant (monotonicity plus the existence of a zero ensures that it
does not fall off as $r\rightarrow\infty$) then the boundary is at
$UV=-\infty$%
\iftoomuchdetail
\begin{detail}%
\ or $\{\tilde{U}=\pm \frac{\pi}{2},\tilde{V}\lessgtr 0\} \cup
\{\tilde{V}=\pm\frac{\pi}{2},\tilde{U}\lessgtr 0\}$%
\end{detail}%
\fi
.  These lines give the
diamond regions familiar
from the Schwarzschild solution.

However, if $r^*(r=\infty)$ is finite---that is, if $f(r)$ diverges as
$r\rightarrow\infty$, as for Schwarzschild-AdS---then the boundary of
spacetime is located at $U V = -e^{r^*(r=\infty)}$.
If \hbox{$r^*(r=\infty)=0$}, then these are
\iftoomuchdetail\begin{detail}the \end{detail}\fi%
vertical lines%
\iftoomuchdetail
\begin{detail}%
\ $\tilde{U}=\pm \frac{\pi}{2} + \tilde{V}$ or $\tilde{Z}=\mp \frac{\pi}{4}$%
\end{detail}%
\fi
.  Thus,
the Penrose diagram looks isomorphic to that for AdS$_{d+1}$, except
for the spacelike singularities in the infinite future and past.
Otherwise, spacelike infinite bows in (out) for $r^*(r=\infty)<0$
($r^*(r=\infty)>0$).
\iftoomuchdetail
\begin{detail}%
This is seen from the expression, for the
$r=\infty$ curves,
\begin{equation}
\tan^2 \tilde{Z} = \frac{e^{r^*(r=\infty)} + \tan^2 \tilde{T}}{
   1+e^{r^*(r=\infty)} \tan^2 \tilde{T}}.
\end{equation}
This is smooth in $r^*(r=\infty)$.
For large $r^*(r=\infty)$, we approach diamonds, which are clearly the
extreme limit of bowing out.  For very small $r^*(r=\infty)$, we
approach the curves for the horizons, which are clearly the extreme
limit of bowing in.
\end{detail}%
\fi
For $r^*(r=\infty)=0$, this gives
$\tan^2 \tilde{Z}=0$, which are the vertical lines of AdS.
The authors of~\cite{fhks} recommend rescaling $U$ and $V$
in the definitions of $\tilde{U}$
and $\tilde{V}$ for the case of finite $r^*(r=\infty)$, so that
spatial infinity is always vertical, and it is the singularities that bow.
However, we have drawn the Penrose diagrams in figure~\ref{fig:PD}
such that the exterior regions bow.
\end{detail}%
\fi

The exterior metric~\eqref{extds} clearly has spherical and
time-translation symmetry.  These clearly extend to the entire
spacetime, although the isometry corresponding to time translation
is null on the horizons, and spacelike inside the black and white holes.
Explicitly,
\begin{equation} \label{time}
\frac{\p}{\p t} = \frac{f'(r_+)}{2} \left( - U \frac{\p}{\p U}
   + V \frac{\p}{\p V} \right)
 = \frac{f'(r_+)}{2}  \left( T \frac{\p}{\p Z}
   + Z \frac{\p}{\p T} \right)\iftoomuchdetail,\else.\fi
\end{equation}
\iftoomuchdetail
\begin{detail}%
or, on the Penrose diagram,
\begin{equation}
\frac{\p}{\p t} = \frac{f'(r_+)}{4} 
  \left( -\sin 2\tilde{U} \frac{\p}{\p \tilde{U}}
         +\sin 2\tilde{V} \frac{\p}{\p \tilde{V}} \right).
\end{equation}
In particular, the norm of the Killing vector is
\begin{equation}
\norm{\frac{\p}{\p t}}^2
= 2 f(r) e^{-f'(r_+) r^*} U V
= 2 f(r) e^{-f'(r_+) r^*} \tan \tilde{U} \tan
\tilde{V}.
\end{equation}
As the factor of $f(r) e^{-f'(r_+) r^*}$ is positive semi-definite ($r^*$
has an imaginary part where for $r<r_+$, where $f(r)<0$), and the
horizons are located at $UV=0$ with the singularities at $UV=1$, we
see that the Killing vector is timelike in the exterior regions and
spacelike inside the horizons.  Finally, we see that, in the
right-hand (left-hand) exterior region the Killing vector is future
(past) directed [since $Z>0$ ($Z<0$)]%
\footnote{On the Penrose diagram, of course, the condition is
$\tilde{Z}\gtrless 0$, but $\tilde{Z} = \frac{1}{2}\left(\tan^{-1}
\frac{T+Z}{2} - \tan^{-1}\frac{T-Z}{2}\right)$ vanishes at $Z=0$ and
is monotonically increasing in $Z$.  In other words $\tilde{Z}\gtrless
0 \Leftrightarrow Z \gtrless 0$.}
and inside the future (past) horizon,
the Killing vector is directed to the right (left).
The Killing vector vanishes at the
intersection point at the center of the diagram.
\end{detail}%
\fi

\section{$\alpha$-Vacua} \label{sec:alpha}

\subsection{Modes}

Recall the definitions~\eqref{RT}, \eqref{RZ},~\eqref{P}.
Although
$R_T$ and $R_Z$ are symmetries of the
spacetime---note that $r^*$ and therefore $r$ is invariant under each
map---%
neither preserves the time-like Killing vector $\frac{\p}{\p t}$.
The combination $R_Z R_T$
preserves all the symmetries of the spacetime---at
least for generic $f(r)$---but has a fixed point at $T=Z=0$.  
The product $P R_Z R_T$ acts freely and preserves all the (generic)
symmetries of the spacetime.  We refer to this as the antipodal map,
${\mathcal P}_A$,
\begin{equation}
{\mathcal P}_A =P R_Z R_T.
\end{equation}
Observe that the antipodal map connects points in opposite exterior
regimes---that is, points for which time has opposite orientation.
The antipodal point of $x=(T,Z,\Omega)$ is denoted
$x_A=(-T,-Z,P \Omega)$.

The wave equation $(-\Box+\mu^2)\phi=0$
for a ``free'' scalar field $\phi$ of mass $\mu$ reads%
\footnote{For simplicity, we assume either no coupling of the field to
the scalar curvature, or else that such a coupling can be made, and
is, part of the definition of $\mu$ to the linearized level.  The
latter assumption is equivalent to demanding that the spacetime have
constant scalar curvature, which, of course, restricts $f$, but is
consistent with demanding a vacuum solution of Einstein's equations
with or without cosmological constant.}
\begin{equation} \label{eom}
\frac{U V f'(r_+)^2}{8 f(r) r^{d-1}} \left[
    \frac{\p}{\p U} r^{d-1} \frac{\p}{\p V} 
  + \frac{\p}{\p V} r^{d-1} \frac{\p}{\p U} \right] \phi
- \frac{1}{r^2} \nabla^2_{S^{d-1}} \phi + \mu^2 \phi = 0,
\qquad r=r(UV).
\end{equation}
This is obviously solved by expanding in spherical harmonics
$Y_{n,s^{(n)}}(\Omega)$ on the
$S^{d-1}$ for which
\begin{equation}
\nabla^2_{S^{d-1}}Y_{n,s^{(n)}}(\Omega) = -n(n+d-2) Y_{n,s^{(n)}}(\Omega).
\end{equation}
Here $s^{(n)}$ denotes the remaining quantum numbers of the spherical
harmonics.

We choose a standard~\cite{a,bms} nonstandard (not a typo)
basis of spherical harmonics for which
\begin{equation}
Y_{n,s^{(n)}}(P \Omega) = Y_{n,s^{(n)}}(\Omega)^*,
\end{equation}
with the asterisk denoting complex conjugation.  Since the equation of
motion~\eqref{eom} is clearly invariant under complex conjugation and
under the antipodal map%
\iftoomuchdetail
\begin{detail}%
\ (because $r$ is a function only of the product $UV$)%
\end{detail}%
\fi
, we use these spherical harmonics to
choose our modes such that
\begin{equation} \label{phiA}
\phi_{\kappa,n,s^{(n)}}(x_A) = \phi_{\kappa,n,s^{(n)}}(x)^*,
\end{equation}
where $\kappa$ is the remaining quantum number(s) needed to specify
the solution, and is (presumably) related to the frequency of the mode.
The Klein-Gordon inner product implies that if
$\phi_{\kappa,n,s^{(n)}}(x)$ is a ``positive
frequency'' mode then $\phi(x)_{\kappa,n,s^{(n)}}^*$ is ``negative
frequency''.
This is, however,
consistent with the choice~\eqref{phiA} as the antipodal map
includes time reversal.

The mode expansion reads
\begin{equation} \label{phimodes}
\phi(x) = \sum_{\kappa,n,s^{(n)}} \left[
   a^{\vphantom{\dagger}}_{\kappa,n,s^{(n)}} 
  \phi^{\vphantom{*}}_{\kappa,n,s^{(n)}}(x)
 + a^\dagger_{\kappa,n,s^{(n)}} \phi^*_{\kappa,n,s^{(n)}}(x) \right],
\end{equation}
Properly normalized modes under the Klein-Gordon inner-product imply
canonical commutation relations for the operators,
\begin{equation}
\com{a^{\vphantom{dagger}}_{\kappa,n,s^{(n\vphantom{'})}}}{
   a^\dagger_{\kappa',n',s^{\prime(n')}}}
 = \delta_{\kappa,\kappa'} \delta_{n,n'} \delta_{s^{(n)},s^{\prime(n')}},
\end{equation}
where, for $\kappa$ the sum and Kronecker-$\delta$ is understood as a
an integration, possibly with discrete sum, and Dirac-$\delta$,
possibly with an additional Kronecker-$\delta$, if $\kappa$ takes
continuum values, possibly with an additional discrete quantum number.
This is understood without comment in the following.

\subsection{Green Functions}

The vacuum is defined to be annihilated by annihilation operators
$a_{\kappa,n,s^{(n)}}$.  Green functions are then found by evaluating
two-point functions.  The Wightman function is
\begin{equation} \label{defG+}
G_0^{(+)}(x,x') = \bra{0} \phi(x) \phi(x') \ket{0}
 = \sum_{\kappa,n,s^{(n)}} 
   \phi^{\vphantom{*}}_{\kappa,n,s^{(n)}}(x) \phi^*_{\kappa,n,s^{(n)}}(x');
\end{equation}
the Hadamard function is
\begin{equation} \label{defG1}
G_0^{(1)}(x,x') = \bra{0} \anti{\phi(x)}{\phi(x')} \ket{0}
 = G^{(+)}(x,x') + G^{(+)}(x',x);
\end{equation}
the commutator function is
\begin{equation} \label{defD}
i D_0(x,x') =\bra{0} \com{\phi(x)}{\phi(x')} \ket{0}
 = G^{(+)}(x,x') - G^{(+)}(x',x)
 = G^{(+)}(x,x') - G^{(+)}(x,x')^*;
\end{equation}
and the Feynman propagator,
\begin{equation} \label{defGF}
\begin{split}
i G^F_{0}(x,x') &= \bra{0} \TO \left( \phi(x)\phi(x') \right) \ket{0}
 = \Theta(x,x') G^{(+)}_0(x,x') + \Theta(x',x) G^{(+)}_0(x',x)
\\ &= \frac{1}{2} G^{(1)}_0(x,x') + \frac{1}{2} \epsilon(x,x') i D_0(x,x'),
\end{split}
\end{equation}
where $\TO$
\iftoomuchdetail\begin{detail}(not to be confused with Kruskal time)
\end{detail}\fi
 denotes
time ordering and
\begin{align}
\Theta(x,x') &= \begin{cases} 1, &\text{$x$ to the future of $x'$}, \\
  \frac{1}{2}, &\text{$x$, $x'$ spacelike separated}, \\
  0, &\text{otherwise},
\end{cases} &
\epsilon(x,x') &= \begin{cases} 1, &\text{$x$ to the future of $x'$}, \\
 0, &\text{$x$, $x'$ spacelike separated}, \\
 -1, &\text{$x$ to the past of $x'$}.
\end{cases}
\end{align}
In general, however, there is no natural choice
of ``positive frequency'' modes and therefore no natural choice of
vacuum; given a particular decomposition, one
can always perform a Bogoliubov transformation to a new,
nonequivalent, decomposition.  However, there are natural choices in
the following senses.  First, if one demands that the vacuum respect
the symmetries of the spacetime, then this restricts the choices of
vacua.  Indeed, it is well known that the Minkowski vacuum is unique.
For de Sitter, there is a one (complex)-parameter family of symmetric vacua
($\alpha$-vacua), obtained from the Bunch-Davies vacuum by a
$\vk$-independent Bogoliubov transformation, where $\vk$ is the
spatial momentum in planar coordinates.\cite{m,a}  Alternatively, one usually
obtains a unique vacuum by analytic continuation from the Euclidean
spacetime, assuming the latter exists; for de Sitter, this
Hartle-Hawking vacuum coincides
with the Bunch-Davies vacuum.

In Ref.~\cite{hs}, it is argued that it is the Boulware vacuum, and
not the Hartle-Hawking vacuum, that is relevant to AdS/CFT for black
hole backgrounds.  However, it should be clear that most of our
discussion does not depend on the details of the vacuum with which we
start.  One
can also construct $\alpha$-vacua starting with the Boulware vacuum,
and our discussion will follow more-or-less identically.

Note that $T\rightarrow i T_E$
in~\eqref{Kds} defines a
perfectly well-defined Euclidean spacetime and $r\geq r_+$.  $r=r_+$
is the origin $T_E=Z=0$.  Alternatively, $t\rightarrow i t_E$ also
gives a well-defined Euclidean spacetime with $r \geq r_+$.  In
particular, observe that $U$ and $V$ are periodic as 
$t\rightarrow t+\frac{4 \pi i}{f'(r_+)}$; this suggests that an
observer in the original $(t,r)$-coordinates sees a temperature 
$\frac{1}{\beta}=\frac{f'(r_+)}{4\pi}$.

It seems difficult to
analyze the symmetries preserved by a general Bogoliubov
transformation without having explicit formulas for modes.  However,
if we can write down Green functions solely in terms of geodesic
distances, then we have certainly found an invariant vacuum.
Our strategy is therefore to study the Green function following Hartle
and Hawking~\cite{hh}.  This will not give us precise detailed
information, but will give us enough analytic structure to study
generic properties.

Once we have the Hartle-Hawking Green function, then we can define
$\alpha$-vacua following~\cite{a}.  In terms of modes, we define
\begin{equation} \label{Bmodes}
\tilde{\phi}_{\kappa,n,s^{(n)}}(x) 
= \cosh \alpha \, \phi{\vphantom{*}}_{\kappa,n,s^{(n)}}(x)
  + e^{i\gamma} \sinh \alpha \, \phi^*_{\kappa,n,s^{(n)}}(x),
\end{equation}
where $\alpha$ and $\gamma$ are real.%
\footnote{Usually one uses $\beta$ rather than $\gamma$, but that
  would confuse the phase with the inverse temperature.
An alternate convention is 
$\tilde{\phi}_{\kappa,n,s^{(n)}}(x) 
{=} N_\alpha \left( \phi^{\vphantom{*}}_{\kappa,n,s^{(n)}}(x)
  + e^\alpha \phi^*_{\kappa,n,s^{(n)}}(x) \right)$, where 
$\real \alpha{<}0$ and $N_\alpha {=} 1/\sqrt{1-e^{\alpha+\alpha^*}}$.
In other words,
$e^{\alpha_{\text{alternate}}} = e^{i \gamma} \tanh \alpha_{\text{here}}$.
\label{ft:altconv}}
This defines new annihilation and creation operators
\begin{gather}
\phi(x) = \sum_{\kappa,n,s^{(n)}} \left[
   \tilde{a}^{\vphantom{\dagger}}_{\kappa,n,s^{(n)}} 
  \tilde{\phi}^{\vphantom{*}}_{\kappa,n,s^{(n)}}
 + \tilde{a}^\dagger_{\kappa,n,s^{(n)}} \tilde{\phi}^*_{\kappa,n,s^{(n)}} 
\right],
\\
\begin{aligned}
\tilde{a}^{\vphantom{\dagger}}_{\kappa,n,s^{(n)}} &= 
\cosh \alpha \, a^{\vphantom{\dagger}}_{\kappa,n,s^{(n)}} 
- e^{-i \gamma} \sinh \alpha \, a^{\dagger}_{\kappa,n,s^{(n)}}, \\
\tilde{a}^{\dagger}_{\kappa,n,s^{(n)}} &= 
\cosh \alpha \, a^{\dagger}_{\kappa,n,s^{(n)}} 
- e^{i \gamma} \sinh \alpha \, a^{\vphantom{\dagger}}_{\kappa,n,s^{(n)}}.
\end{aligned}
\end{gather}
The vacuum annihilated by $\tilde{a}_{\kappa,n,s^{(n)}}$ is the
$\alpha$-vacuum $\ket{\alpha\gamma}$.  It preserves the symmetries of the
spacetime if $\ket{0}$ does.  Specifically, because of the
choice~\eqref{phiA}, we have, for the Wightman function for example,
\begin{multline} \label{G+ab}
G^{(+)}_{\alpha\gamma}(x,x') 
= \bra{\alpha\gamma} \phi(x) \phi(x') \ket{\alpha\gamma}
= \cosh^2 \alpha \, G_0^{(+)}(x,x')
 + e^{i \gamma} \sinh \alpha \cosh \alpha \, G_0^{(+)}(x_A,x') \\
 + e^{-i \gamma} \sinh \alpha \cosh \alpha \, G_0^{(+)}(x,x'_A)
 + \sinh^2 \alpha \, G_0^{(+)}(x_A,x'_A).
\end{multline}
Since the antipodal map commutes with the symmetries of the spacetime,
and since $\ket{0}$ was assumed to preserve the symmetries, thereby
implying that $G_0^{(+)}(x,x')$ repects the symmetries, it follows that
$G^{(+)}_{\alpha\gamma}(x,x')$ respects the symmetries of the spacetime.
Also, note that
\begin{equation}
G^{(+)}_0(x_A,x'_A) = G^{(+)}_0(x',x) = G^{(+)}_0(x,x')^*.
\end{equation}
Therefore,
\begin{equation}
\begin{split}
G^{(1)}_{\alpha\gamma}&(x,x') 
= G^{(+)}_{\alpha\gamma}(x,x') + G^{(+)}_{\alpha\gamma}(x',x), \\
&= \left(\cosh^2 \alpha+\sinh^2\alpha\right) G_0^{(1)}(x,x')
 + e^{i\gamma} \sinh 2 \alpha \, G_0^{(+)}(x_A,x')
 + e^{-i\gamma} \sinh 2 \alpha \, G_0^{(+)}(x,x'_A), \\
&= \cosh 2 \alpha \, G_0^{(1)}(x,x') 
+ \cos \gamma \sinh 2\alpha \, G_0^{(1)}(x_A,x')
- \sin \gamma \sinh 2\alpha \, D_0(x_A,x').
\end{split}
\end{equation}
Similarly,
\begin{equation}
i D_{\alpha\gamma}(x,x') 
= G^{(+)}_{\alpha\gamma}(x,x') - G^{(+)}_{\alpha\gamma}(x',x)
= i D_0(x,x');
\end{equation}
this also follows since the commutator of the field is a c-number,
independent of the vacuum.  Finally,
\begin{equation} \label{GFab}
\begin{split}
i G^F_{\alpha\gamma}(x,x')
&= \frac{1}{2} G_{\alpha\gamma}^{(1)}(x,x')
  + \frac{1}{2} \epsilon(x,x') i D_{\alpha\gamma}(x,x'), \\
&= i G^F_0(x,x') + \frac{1}{2} \left[G_{\alpha\gamma}^{(1)}(x,x')
    - G_0^{(1)}(x,x') \right],
\end{split}
\end{equation}
since $D_{\alpha\gamma}=D_0$ can be rewritten in terms of the Feynman
propagator.  These formulas---and derivations---were first given by
Allen~\cite{a} for
de Sitter.  In particular, the Green functions of the two vacua differ
only by a homogeneous solution of the wave equation.
Alternatively,
\begin{multline} \label{GFalpha}
i G^F_{\alpha\gamma}(x,x')
= \cosh^2 \alpha \, i G_0^F(x,x') + \sinh^2 \alpha \, i G_0^F(x,x')^*
\\  + \cos \gamma \sinh 2 \alpha \, G^{(1)}_0(x_A,x')
  - \sin \gamma \sinh 2 \alpha \, D_0(x_A,x').
\end{multline}
The $\alpha$-vacuum Feynman propagator is written in terms of Green
functions
beyond 
just the Hartle-Hawking Feynman propagator.  Also one term
involves a complex conjugate, which implies the opposite
time ordering.\cite{el,el2}
(More precisely, this is the Hartle-Hawking Feynman propagator
evaluated at the antipodal points, but with opposite time ordering
from that of the antipodal points.)

When $\sin \gamma\neq 0$, the Feynman propagator involves the
commutator function.  As explained in~\cite{a},
the commutator function is antisymmetric in $x,x'$ whereas the
geodesic distance is symmetric in $x,x'$.  In particular, the
commutator function depends on the time ordering and is therefore not
invariant under CPT.  So the $\alpha$-vacua are only CPT-invariant for
$\gamma=0,\pi$.\cite{a}

\subsection{Example: The BTZ Black Hole}

The three-dimensional BTZ black hole\cite{BTZ,BTZ2} is given by
\begin{equation}
ds^2 = -(\frac{r^2}{\ell^2} - M) dt^2 + \frac{dr^2}{\frac{r^2}{\ell^2} - M}
   + r^2 d\phi^2,
\end{equation}
which is the case of $f(r)=\frac{r^2}{\ell^2} - M$; thus
$r_+=\sqrt{M} \Lambda$.
In Kruskal coordinates,
\begin{equation}
ds^2 = -4 \frac{r_+^2}{M} (1+UV)^{-2} dU dV
   + r_+^2 \frac{(1-UV)^2}{(1+UV)^2} d\phi^2.
\end{equation}
Spatial infinity is $UV=-1$ and the singularity is at $UV=1$.

It is well-known that the BTZ black hole is a quotient of AdS$_3$.\cite{BTZ2}
That is, with AdS$_3$ embedding coordinates $X$,
\begin{equation}
-(X^{-1})^2 - (X^0)^2 + (X^1)^2 + (X^2)^2 = -\ell^2,
\end{equation}
we can take
\begin{equation}
\begin{aligned}
X^{-1} &= \frac{r}{\sqrt{M}} \cosh \sqrt{M}\phi, & \qquad
X^0 &= \frac{\ell}{\sqrt{M}} \sqrt{\frac{r^2}{\ell^2}-M}
  \sinh \frac{\sqrt{M}}{\ell} t, \\
X^1 &= \frac{r}{\sqrt{M}} \sinh \sqrt{M}\phi, & \qquad
X^2 &= \frac{\ell}{\sqrt{M}} \sqrt{\frac{r^2}{\ell^2}-M}
  \cosh \frac{\sqrt{M}}{\ell} t,
\end{aligned}
\end{equation}
or, for Kruskal coordinates,
\begin{equation}
\begin{aligned}
X^{-1} &= \ell \frac{1-UV}{1+UV} \cosh \sqrt{M}\phi, & \qquad
X^0 &= \ell \frac{U+V}{1+UV} \\
X^1 &= \ell \frac{1-UV}{1+UV} \sinh \sqrt{M}\phi, & \qquad
X^2 &= \ell \frac{U-V}{1+UV}.
\end{aligned}
\end{equation}
With an infinite range for $\phi$, these describe AdS$_3$; the BTZ
black hole is obtained by identifying $\phi\sim \phi+2\pi$.

The Green function for a scalar field 
for the BTZ black hole is obtained by the method of images
(see, for example,~\cite{shir,is,mz,esko,hkk,kos}).
For two particles in the same
exterior region, with particle 2 on the boundary ($r_2=\infty$), it reads
\begin{equation} \label{BTZG}
G(X_1,X_2) = \sum_{n=-\infty}^\infty
 \left[-\sqrt{\frac{r_1^2-r_+^2}{r_+^2} 
      \cosh(\frac{r_+ \Delta t}{\ell^2})
  + \frac{r_1}{r_+} \cosh \frac{r_+}{\ell} (\Delta \phi + 2\pi n)} \;
  \right]^{-2 h_+},
\end{equation}
where $2 h_+ = 1 + \sqrt{1+\ell^2 m^2}$.
The authors of~\cite{hkk,kos} noted that this is well-behaved under
analytic continuation to $X_1$ in the other exterior region via
$t\rightarrow t-i \frac{\pi \ell^2}{r_+}$; this continues when one
additionally takes $\phi\rightarrow \phi+\pi$.
That is, the Green function is well-behaved if $X_1$ is replaced with
$X_{1A}$.

Alternatively, one might note that even pure AdS has a natural
antipodal map, defined in the embedding space by
\begin{equation}
(X^{-1},X^0,X^1,X^2) \longrightarrow (-X^{-1},-X^0,-X^1,-X^2),
\end{equation}
and therefore in the BTZ coordinates by
\begin{equation} \label{wrongmap}
t\longrightarrow t-i \frac{\pi \ell^2}{r_+}, \qquad
\phi\longrightarrow \phi +
\frac{\pi}{\sqrt{M}} i.
\end{equation}
Actually, this map is clearer in AdS global global coordinates, for
which
\begin{equation} \label{globalAdS}
\begin{aligned}
X^{-1} &= \ell \cosh \xi \sin \tau, & X^1 &= \ell \sinh \xi \cos\theta, \\
X^0 &= \ell \cosh \xi \cos \tau, &    X^2 &= \ell \sinh \xi \sin \theta,
\end{aligned}
\end{equation}
and so the map is just
\begin{equation} \label{wrongmapglobal}
\tau \longrightarrow \tau + \pi, \qquad \theta\longrightarrow \theta+\pi,
\end{equation}
From the point of view of $\alpha$-vacua, this is somewhat peculiar;
this produces timelike
correlations; see Fig.~\ref{fig:AdS3ids}.
(Also, for AdS, rather than the BTZ black hole,
if we are avoiding closed timelike curves by 
working on the covering space of AdS, this map is isomorphic to
$\ZZ$, not $\ZZ_2$, and so produces an infinite number of correlations.)
More to the point, however, we note that, unlike the antipodal map 
\hbox{$(t,\phi)\rightarrow (t-i\frac{\pi \ell^2}{r_+},\phi+\pi)$},
the map~\eqref{wrongmap} only changes the
Green function~\eqref{BTZG} by a phase, and so does not accomplish anything!
Though we have only demonstrated this for the bulk-to-boundary
propagator, this is in fact true for the bulk-to-bulk
propagator~\cite{kos,bl,io}.
\iftoomuchdetail
\begin{detail}%
\begingroup \small
[Proof: From~\cite[Eq. (14)]{io},
say, the AdS$_{d+1}$ propagator for a mass-squared $m^2$ 
scalar field in global coordinates is
\begin{equation}
\begin{aligned}
G^F(\tau_1,\xi_1;\tau_2=0,\xi_2=0;m^2)
&= -N \sigma(\tau_1,\xi_1)^{-2\nu}
  {_2F_1}\left(\nu,\nu+\thalf;2\nu+1-\tfrac{d}{2};
     \sigma(\tau_1,\xi_1)^{-2}-i\epsilon\right), \\
N &\equiv -2^{-2\nu-1} \pi^{-d/2} \ell^{1-d} 
  \frac{\Gamma(2\nu)}{2\nu+1-\tfrac{d}{2}}, \\
\nu &\equiv \tfrac{d}{4} + \thalf \sqrt{\tfrac{d^2}{4} + m^2 \ell^2}, \\
\sigma(\tau_1,\xi_1) & \equiv \cos\tau_1 \cosh\xi_1
\end{aligned}
\end{equation}
(As per footnote~\ref{ft:AdSO+O-} (p.~\pageref{ft:AdSO+O-}), there is a
range of mass-squareds for which one could also choose the other
square-root for $\nu$.)
Note that the choice for the second point is without loss of
generality, by SO$(2,d-2)$-invariance---that is, by the choice of
$X$'s in Eq.~\eqref{globalAdS}.  The propagator between point 2 and
the antipodal of point 1 is obtained by taking $\tau_1\longrightarrow
\tau_1+\pi$; this just changes the sign of $\sigma$, thereby
contributing a phase $(-1)^{-2\nu}$, but not affecting the argument of
the hypergeometric function.]
\endgroup
\end{detail}%
\fi
This
is in accord with Witten's demonstration~\cite{ew}
that the pure AdS Green function
is essentially unique.

\FIGURE[t]{
\includegraphics[height=2.5in,clip=true,keepaspectratio=true]{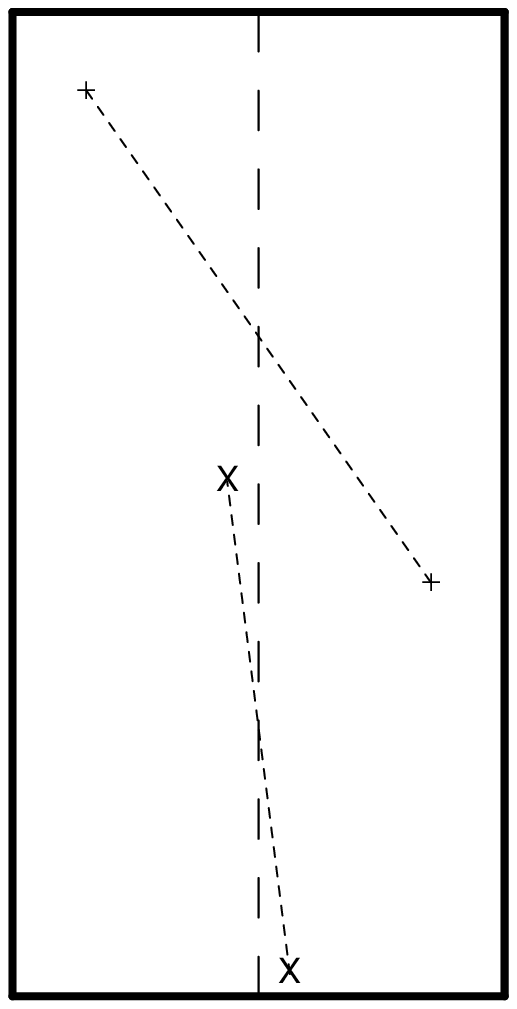}
\caption{The Penrose diagram for AdS$_3$, including both $\theta=\pi$
(left-hand side of the diagram) and $\theta=0$ (the right-hand side).
The dashed line is the (spatial) origin of AdS$_3$ ($\xi=0$ in
Eq.~\eqref{globalAdS}).
This is AdS$_3$, and not its cover.
We have drawn two pairs of points that are mapped to each other under
the natural
AdS$_3$ antipodal map.  This antipodal map correlates
timelike-related points, and is not the one we use to
construct $\alpha$-vacua.%
\label{fig:AdS3ids}}
}

\subsection{The Hartle-Hawking Vacuum}

\subsubsection{Propagators}

We now attempt to understand the Feynman propagator in the
Hartle-Hawking vacuum, following the original paper~\cite{hh}.
The idea is to look at $G^F_0(x,x')$ on the complexified spacetime
($T, Z$ complex) with $x'$ exterior (in region I) to the black
hole and $x$ on the future horizon.
General considerations, which we will not review here%
\footnote{Except we note that
it is important that the complexified spacetime is
nonsingular and complete outside the horizon $r=r_+$.  Then
the path integral be made well-defined via analytic continuation to
the Euclidean regime.  We have already checked this.  The remaining
details appear to be independent of the spacetime in question.}
imply that, in
the complexified spacetime, the
propagator is an analytic function of $x$ and $x'$, except for
singularities displaced slightly from null geodesics.  Hartle and
Hawking write~\cite{hh}
\begin{equation} \label{HHG}
G^F_0(x,x') = K_0(x,x') 
- i \sum_c \frac{e^{i s_c(x,x')/4 W_0}}{s_c(x,x')+i \epsilon} D_c(x,x'),
\end{equation}
where $c$ labels the (complex) geodesics connecting $x$ and $x'$,
$s_c(x,x')$ is the geodesic distance along the $c$th geodesic, $K_0$
and $D_c$ are analytic functions, $W_0$ is a small value of the
Schwinger parameter, and $\epsilon>0$.  We see that the singularities occur for
geodesics that are slightly displaced from null ones.
\iftoomuchdetail
\begin{detail}%
Therefore we study null
geodesics.

Without loss of generality, null geodesics are taken to be in the
``equatorial plane''; in terms of standard polar coordinates, this is
$\theta_1=\dots=\theta_{d-2}=0$ and $\phi\in [0,2\pi)$.  Working in
$(v,r)$ coordinates---recall $v=t+r^*$---the metric
\begin{equation}
ds^2 = -f(r) dv^2 + 2 dr dv + 
r^2 \left(d\theta_1^2 + \sin^2\theta_1 \left( \skipthis{d\theta_1^2 +} \dots
\left( d\theta_{d-2}^2 + \sin^2 \theta_{d-2} d\phi^2 \right) \dots \right)
\right),
\end{equation}
is nonsingular at the horizon (nevertheless, for real $v$ these coordinates
only cover the $V>0$ patch of the spacetime).
Clearly $\frac{\p}{\p v}=\frac{\p}{\p t}$ is  Killing,
as is $\frac{\p}{\p \phi}$.
In the equatorial plane, the corresponding conserved quantities are
\begin{align}
e &= f(r) \frac{dv}{d\lambda} - \frac{dr}{d\lambda}, &
\ell &= r^2 \frac{d\phi}{d\lambda},
\end{align}
where $\lambda$ is the affine parameter of the null geodesic.
Since we consider the complexified spacetime, we allow $e$ and $\ell$
to be complex.  Note, however, that for real geodesics,
$\frac{dr}{d\lambda}<0$ (ingoing) and $\frac{dv}{d\lambda}>0$ (future
directed) and so $e>0$.
That
the geodesic is null implies that
\begin{equation}
-f(r) \left(\frac{dv}{d\lambda}\right)^2 
+ 2 \frac{dr}{d\lambda} \frac{dv}{d\lambda}
+ r^2 \left(\frac{d\phi}{d\lambda}\right)^2 = 0.
\end{equation}
This can be solved for $\frac{dr}{d\lambda}$ in terms of $e$ and
$\ell$; then the position $x$ at which the null geodesic originating
at the exterior point $x'$ crosses the future horizon ($r=r_+$) is given by
\begin{equation} \label{vandphi}
\begin{aligned}
v-v' &= \int_{r'}^{r_+} \frac{dr}{f(r)}
  \left[1-\frac{1}{\sqrt{1-\frac{b^2}{r^2}f(r)}}\right], \\
\phi-\phi' &= -b \int_{r'}^{r_+} \frac{dr}{r^2 \sqrt{1-\frac{b^2}{r^2}f(r)}},
\end{aligned}
\end{equation}
where $b\equiv \frac{\ell}{e}$ is the impact parameter.
(Note that $\phi-\phi'>0$ for positive $\ell$ since $r_+<r'$.)
In
appropriate dimensionless units, there is a range
$0<b\leq b_0$ below which null geodesics cross the 
future horizon for real values of
$v,\phi$, provided $f(r)$ does not grow faster than $r^2$.  (This
upper bound holds for Schwarzschild-AdS, for which $f(r)\sim r^2$, and
for which $b_0$ has a complicated expression which, for large black
hole mass, can be approximated by $b_0 \approx \Lambda$.)  
Therefore, assuming $f(r)$ does not grow faster than $r^2$, we have
the same situation as for Schwarzschild~\cite{hh}, namely 
complex null geodesics which cross the future horizon with real values
of $\phi$ also have real values of $v$ and $V$.  A similar statement
for the past horizon is obvious%
\iftoomuchdetail
\begin{detail}%
; one simply changes the sign of the square roots in the
integrals~\eqref{vandphi}; this corresponds to $e<0$,
which follows from the definition of $e$, the requirement
that the geodesic be past directed ($\frac{dv}{d\lambda}<0$) and
the null condition, which implies
\begin{equation}
\frac{dr}{d\lambda} = \frac{1}{2} f(r) \frac{dv}{d\lambda}
   - \frac{r^2}{2} \left.\left(\frac{d\phi}{d\lambda}\right)^2\right/
        \frac{dv}{d\lambda}>\frac{1}{2}f(r) \frac{dv}{d\lambda}.
\end{equation}%
\end{detail}%
\else
.
\fi

Now t%
\end{detail}%
\else

T%
\fi
he question is how $x$ should be perturbed, in the complex plane,
in order to sit on the singularity in the propagator~\eqref{HHG}.  The
general analysis is precisely as in~\cite{hh}.
\iftoomuchdetail
\begin{detail}%
If $x_0$ is the point at which a null geodesic from $x'$ hits the
horizon, then if $x$ is near $x_0$,
\begin{equation}
s(x,x') = \evalat{\frac{\p s(x,x')}{\p V}}{V=V_0}(V-V_0) + \order{(V-V_0)^2}.
\end{equation}
We want to know the direction in which this is negative-imaginary;
this, of course, depends on the sign of 
$\evalat{\frac{\p s(x,x')}{\p V}}{V=V_0}$.  Let the tangent to the
null geodesic ending at $x_0$ be denoted by $\xi$.  Perturbing away from $V_0$
in the positive $V$ direction---that is, in the forward null direction
parallel to the horizon---will bring us to the end point of a
neighbouring, timelike, geodesic which started from $x'$.  Let the
displacement between the two geodesics be denoted by $\zeta$.  By
definition, on the horizon, $\zeta=\frac{\p}{\p V}$; since $\xi$ and
$\zeta$ are both future pointing and null at the horizon, but are not
parallel, clearly $\xi\cdot\zeta <0$.  The equation of geodesic
deviation implies that $\frac{d^2 \xi \cdot\zeta}{d\lambda^2}=0$.
Thus, as the two geodesics start out at $x'$,
$\xi\cdot\zeta \propto \lambda$, where $\lambda=0$ labels the start of the
geodesics (the point $x'$), with negative coefficent (since the value
is negative at the horizon).  In particular, the tangent vector to the
new geodesic, namely $\xi + \zeta \delta V$, is timelike.
This is a long-winded way of proving what we started by saying,
that the geodesic connecting $x'$ with $x=x_0+\zeta \delta V$ is
timelike.
Thus, $\evalat{\frac{\p s(x,x')}{\p V}}{V=V_0}$ is negative and so the
\end{detail}%
\else
The
\fi
singularities of the propagator are in the upper-half $V$ plane%
\iftoomuchdetail
\begin{detail}%
.

Similarly, since the past horizon is at $V=0$, and $x'$ and
$x_0+\delta U \frac{\p}{\p U}$ are spacelike related (for positive
$\delta U$), we find that the singularities of the propagator are in
\end{detail}%
\else
\ and
\fi
the lower-half $U$ plane.  
For completeness, we note that the reverse is true for the other
exterior region, since the roles of $U$ and $V$ are reversed there.

That the singularities are located in opposite sides of the real axis
of the $U$ and $V$ planes implies that the singularities are 
in the complex $t$-plane, and not in the complex $r$-plane.
Specifically, the singularities are located to the future of $t'$
and just above the real axis
of the complex $t$-plane, as well as at the image points of these under
the $\beta i = \frac{4\pi i}{f'(r_+)}$ periodicity.  Moreover
since 
$t+\frac{\beta}{2} i$ corresponds to the antipodal point, the singularity
structure is reversed there.  In other words, there are also
singularities in the past of $t'$ and
just below the line $t=\frac{\beta}{2} i$, and its
periodic images.  This is depicted in Fig.~\ref{fig:Gsing}\ref{fig:Gext}.

\FIGURE[t]{
\begin{tabular}{cc}
\subfig[$x$ in region II or III]{%
    \includegraphics[width=3in,height=3.25in,clip=true,%
              keepaspectratio=true]{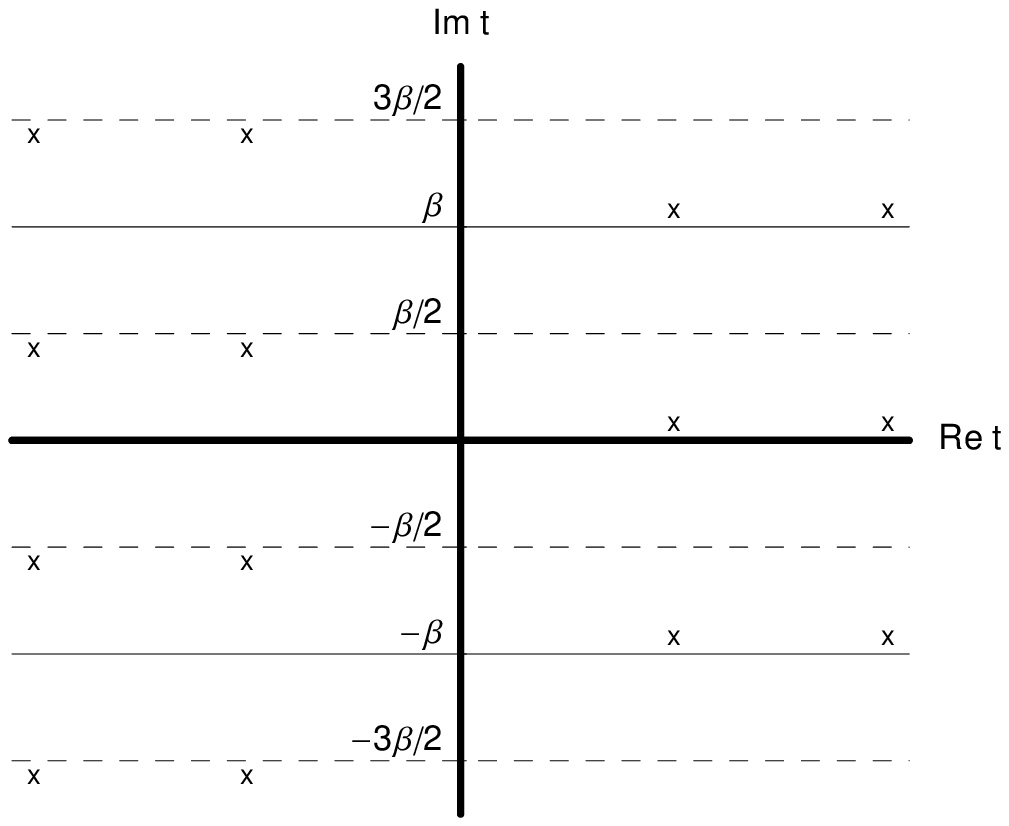}}
    \label{fig:Gext} &
\subfig[$x$ in region I]{%
    \includegraphics[width=3in,height=3.25in,clip=true,%
              keepaspectratio=true]{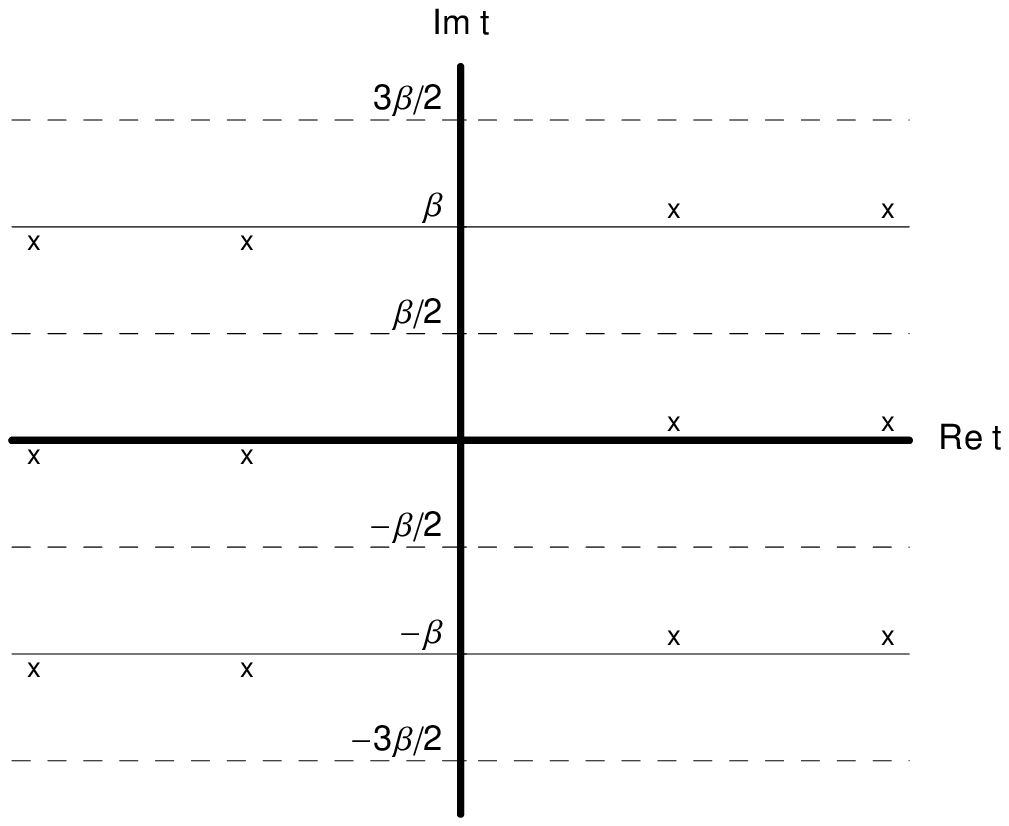}}
    \label{fig:Gextext}
\end{tabular}
\caption{The singularity structure of $G^F_0(x,x')$ when $x'$ is in
region I.  For $x'$ in region IV, the figures
are shifted vertically by $\frac{\beta}{2}$.%
\label{fig:Gsing}}
}

If $x$ is in the same exterior region as $x'$, then there are real null
geodesics for $x$ both in the past of $x'$ as well as the future of
$x'$.  Therefore, the singularities are displaced only from the real
axis (and its images) as in Fig.~\ref{fig:Gsing}\ref{fig:Gextext}.
This is the same singularity structure as for Minkowski space.
There are no singularities near $\im t =  \frac{\beta}{2}$,
which corresponds to the opposite
exterior region and therefore corresponds to points that are
spacelike separated from $x'$.

\iftoomuchdetail
\begin{detail}%
It might appear that these singularities are actually branch cuts,
since the real value of $t$---provided it is bigger (less) than $t'$
for points on the future (past) horizon---can be varied continuously
by varying $b$.  But, of course, this will generically change $\phi$
as well, and we are interested in the singularities of the Green
function in $x$.  Keeping $\phi$ fixed quantizes possible values of $b$
(since $\phi$ and $\phi+2\pi$ are equivalent) and so the singularities
are discrete.  This of course, is essential to having a well-defined
Green function.
\end{detail}%
\fi

\FIGURE[t]{
\begin{tabular}{cc}
\subfig[$x$ in region II or III]{%
    \includegraphics[width=3in,height=3.25in,clip=true,%
              keepaspectratio=true]{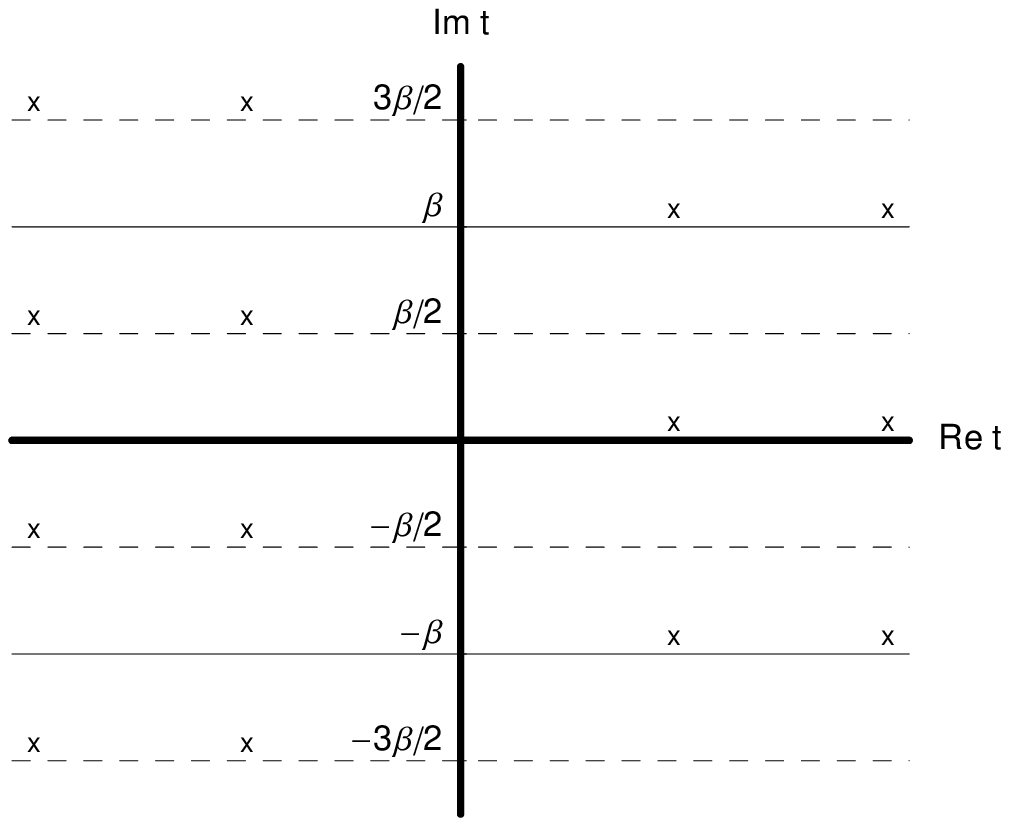}}
    \label{fig:Wext} &
\subfig[$x$ in region I]{%
    \includegraphics[width=3in,height=3.25in,clip=true,%
              keepaspectratio=true]{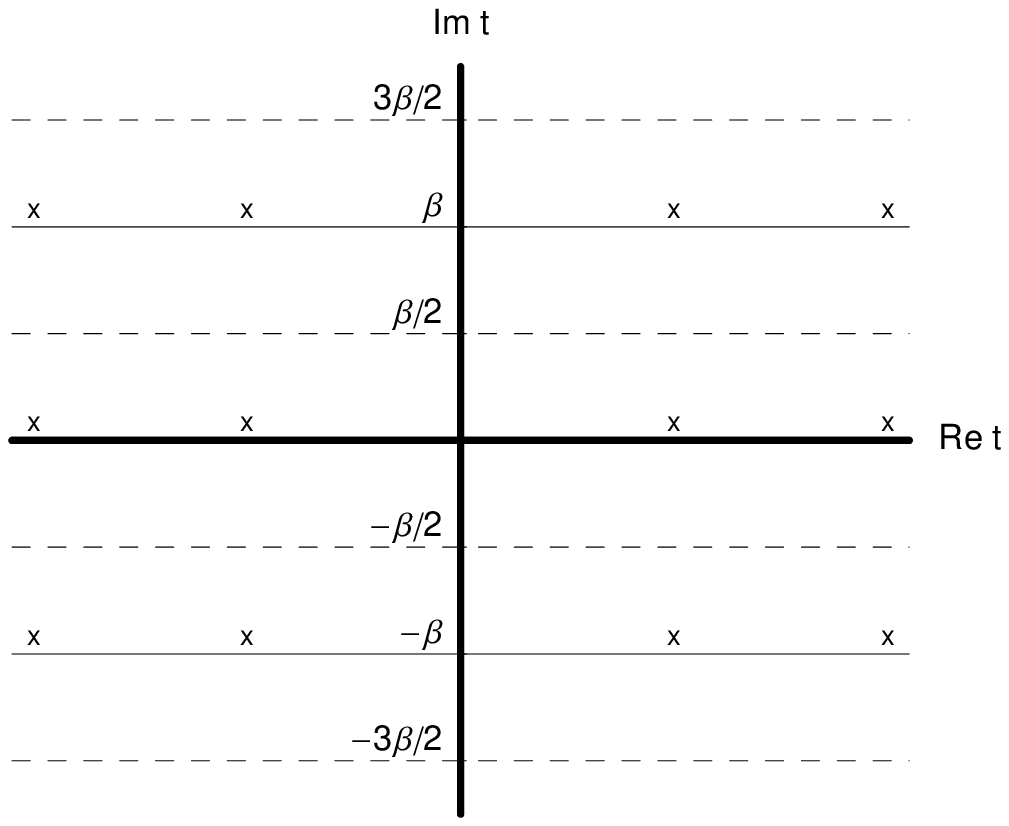}}
    \label{fig:Wextext}
\end{tabular}
\caption{The singularity structure of $G^{(+)}_0(x,x')$ when $x'$ is in
region I.  For $x'$ in region IV, the figures
are shifted vertically by $\frac{\beta}{2}$.%
\label{fig:Wsing}}
}

Now that we know the singularity structure of the Feynman propagator,
we can, from~\eqref{defGF}, deduce the singularity structure of the
Wightman function~\eqref{defG+}.  More precisely, since the Feynman
propagator is time ordered, the singularity
structure of the Wightman function and the Green function agree for
$\real t>\real t'$---at least when this is a sensible time ordering.
For the opposite regime of $\real t$, we can observe that
\begin{equation} \label{exchgW}
G_0^{(+)}(x,x') = G_0^{(+)}(x',x)^*,
\end{equation}
and so we know the singularity structure.  It is depicted in
Fig.~\ref{fig:Wsing}.  Note that, unlike the situation for the Feynman
propagator, if $x'$ is exterior and $x$ is interior it is impossible
to shift $t$ by
$\pm \frac{\beta}{2} i$ without crossing a singularity of the Wightman
function for some $\real t$.

\subsubsection{Thermality\skipthis{ of the Hartle-Hawking Vacuum}}

Given this singularity structure of the Green functions, we can
examine thermality of the spacetime.
The calculation is essentially identical to that in~\cite{bms}.
Explicitly, the transition rate
rate for a
detector which follows a trajectory $x'(\tau)$ ($\tau$ is proper time)
to jump from energy $E_i$
to $E_j$ is given by~\cite{bd}
\begin{equation} \label{P0}
P_0(E_i\rightarrow E_j) = \abs{m_{ij}}^2
    \int_{-\infty}^\infty d\tau e^{-i \Delta \epsilon \,\tau} 
G^{(+)}_0(x(\tau),x(0)),
\qquad \Delta \epsilon \equiv \epsilon_j-\epsilon_i,
\end{equation}
where $\abs{m_{ij}}^2$ is a matrix element that depends on the details of
the detector.
Here $\tau=0$ has been defined to be the proper time at which
$x(0)=x'$, and $\epsilon_{i,j}$ are the proper energies.  We consider
a detector trajectory at a fixed point
exterior to the horizon, $r=R$.  Then $\tau = \sqrt{f(R)} (t-t')$ and
$\Delta \epsilon = \Delta E/\sqrt{f(R)}$.
Since the detector stays in the exterior
region, we only need the singularity structure for
exterior points, depicted in Fig.~\ref{fig:Wsing}\ref{fig:Wextext}.
This singularity structure allows us to shift the contour down to 
$\im t = -\frac{\beta}{2}$; thus,
\begin{subequations}
\begin{align} \label{Porig}
P_0(E_i\rightarrow E_j)
&= \abs{m_{ij}}^2 \sqrt{f(R)}
    \int_{-\infty}^\infty dt\,  e^{-i \Delta E (t-t')} 
    G^{(+)}_0\left(t,R;t',R\right),
\\ \label{shiftP}
&= \abs{m_{ij}}^2 \sqrt{f(R)} e^{-\frac{\beta \Delta E}{2}}
    \int_{-\infty}^\infty dt\,  e^{-i \Delta E (t-t')} 
    G^{(+)}_0\left(t-i\tfrac{\beta}{2},R;t',R\right).
\end{align}
\end{subequations}%

As the point $x$ is in region I, the argument of the Wightman
function in~\eqref{shiftP} corresponds to a point in region IV.  Such
a point is spacelike separated from the point $x'$.  For spacelike
separated points, general principles imply that the commutator
function~\eqref{defD} vanishes, and the Wightman function is
symmetric.  Employing time translation invariance---or more precisely,
that the Wightman function is a function of the geodesic distance---we
can rephrase this statement as
\begin{equation} \label{G+-t}
G^{(+)}_0\left(t-i\tfrac{\beta}{2},R;t',R\right)
= G^{(+)}_0\left(t'-t+i\tfrac{\beta}{2},R;0,R\right),
\end{equation}
since both sides of the equation are a function of the same geodesic
distance, and there are no nearby singularities.  Upon using
eq.~\eqref{G+-t},
$\beta i$ periodicity and replacing the integration
variable $t$ by $t'-t$,
equation~\eqref{shiftP} reads
\begin{equation}
\begin{split}
P_0(E_i\rightarrow E_j)
&= \abs{m_{ij}}^2 \sqrt{f(R)} e^{-\frac{\beta \Delta E}{2}}
    \int_{-\infty}^\infty dt\,  e^{i \Delta E\, t} 
    G^{(+)}_0\left(t-i\tfrac{\beta}{2},R;0,R\right),
\\
&= \abs{m_{ij}}^2 \sqrt{f(R)} e^{-\beta \Delta E}
    \int_{-\infty}^\infty dt\,  e^{i \Delta E\, t}
    G^{(+)}_0\left(t,R;0,R\right),
\end{split}
\end{equation}
upon shifting the contour again.  In other words, using~\eqref{Porig}
on the right-hand side,
\begin{equation} \label{P0rat}
\frac{P_0(E_i\rightarrow E_j)}{P_0(E_j\rightarrow E_i)}
 = e^{-\beta \Delta E}.
\end{equation}
This is the condition
for detailed balance at temperature
\begin{equation} \label{Thh}
T = \beta^{-1}
\end{equation}
as predicted from the periodicity of Euclidean time.

\subsection{Thermality of $\alpha$-Vacua}

Let us repeat the calculation of detailed balance for an
$\alpha$-vacuum.  We again start with
\begin{equation} \label{Palpha}
P_{\alpha\gamma}(E_i\rightarrow E_j)
= \abs{m_{ij}}^2 \sqrt{f(R)}
    \int_{-\infty}^\infty dt\,  e^{-i \Delta E\, (t-t')} 
    G^{(+)}_{\alpha\gamma}(t,R;t',R);
\end{equation}
the difference between~\eqref{P0} and~\eqref{Palpha} is in 
the Wightman function.
Substituting~\eqref{G+ab}, we find
\begin{multline}
P_{\alpha\gamma}(E_i\rightarrow E_j)
= \abs{m_{ij}}^2 \sqrt{f(R)}
    \int_{-\infty}^\infty dt\,  e^{-i \Delta E\, (t-t')} 
\left\{ \cosh^2 \alpha\,  G^{(+)}_0(t,R;t',R) 
\right. \\ \left.
      + \frac{1}{2} e^{i \gamma} \sinh 2 \alpha\,  
             G_0^{(+)}\left(t - i\tfrac{\beta}{2},R;t',R\right)
      + \frac{1}{2} e^{-i \gamma} \sinh 2 \alpha\,  
             G_0^{(+)}\left(t,R;t' + i\tfrac{\beta}{2},R\right)
\right. \\ \left.
      + \sinh^2 \alpha\,  G^{(+)}_0\left(t - i\tfrac{\beta}{2},R;
                                      t'+ i\tfrac{\beta}{2},R\right)
\right\}
\end{multline}
Again, since for spacelike separated points,
the Wightman function depends only on the geodesic
distance, we can use
\begin{equation} \label{st'}
G_0^{(+)}\left(t,R;t' + i\tfrac{\beta}{2},R\right)
= G_0^{(+)}\left(t-i\tfrac{\beta}{2},R;t',R\right),
\end{equation}
for the middle two terms, and appropriately shift the contour.
For the last term, we shift the contour%
\footnote{Here $x'$ is in region IV, so the poles in $t$ will also be
in region IV; {\em i.e.\/} just off the $\im t=-\beta i/2$ line,
and its images.  Moreover, time is reversed in region IV, so the poles
are just below this line---the opposite of
Fig.~\ref{fig:Wsing}\ref{fig:Wextext} in which the poles are just
above $\im t=0$ and its images.  Therefore we can shift the
$t$ contour up to the real axis
without crossing any poles.}
and then use~\eqref{st'} and shift the contour again.  As a result,
\begin{equation}
P_{\alpha\gamma}(E_i\rightarrow E_j)
=  \abs{m_{ij}}^2 \sqrt{f(R)}
\abs{ \cosh \alpha
   + \sinh \alpha e^{i\gamma} e^{\frac{\beta}{2} \Delta E}
}^2 P_0(E_i\rightarrow E_j).
\end{equation}
Thus,
\begin{equation} \label{Prat}
\frac{P_{\alpha\gamma}(E_i\rightarrow E_j)}{
  P_{\alpha\gamma}(E_j\rightarrow E_i)}
= \abs{\frac{ \cosh \alpha
   + \sinh \alpha e^{i\gamma} e^{\frac{\beta}{2} \Delta E}}{
  \cosh \alpha
   + \sinh \alpha e^{i\gamma} e^{-\frac{\beta}{2} \Delta E}}
}^2
e^{-\beta \Delta E}.
\end{equation}
This is not thermal.  After taking into account the
conventions described in footnote~\ref{ft:altconv}
(p.~\pageref{ft:altconv}), it matches
the nonthermal expression in~\cite{bms}.  Indeed, the authors
of~\cite{kklss} have already
observed that the de Sitter result depends only on the
analytic structure of the Green function.

\section{(No) Objections to $\alpha$-Vacua} \label{sec:object}

There has been a substantial amount of work attempting to
debunk the idea of $\alpha$-vacua for de
Sitter space.  In this section we will address the objections and
describe how they may be evaded in black hole spacetimes.

We will see that some of the objections involve consideration of spacetime
points that are separated by
the black hole horizon.  Issues involving causality and correlation functions
of such points are already poorly understood, and are related to questions
involving information loss in, and the end-point of, Hawking radiation.
For example~\cite{bhc,bhc2}, the
concept of black hole complementarity has been suggested as a means to
try to assert the consistency of black hole physics.  According to
black hole complementarity, one may describe a black hole
system as seen by an observer who falls through the horizon and sees no
irregularities as she falls through the horizon, or as seen by
an observer constrained to live outside the horizon and who sees the Hawking
radiation and other thermal physics of the black hole, but one may not
describe the black hole using both observers as these are complentary
descriptions.  This avoids, for example, the quantum Xerox problem
(\eg~\cite{qx})---that black hole evaporation combined with
the causal disconnect between the two sides of the black hole horizon
requires that the quantum information of the infalling matter be copied
to outside states, in conflict with the unitarity of quantum mechanics---but
also leaves open the question of how one should treat quantities such as
correlation functions between points inside and outside the black hole
horizon.

Thus, since physics connecting the interior and exterior of black holes is
poorly understood, even for the well-known vacua, we should not expect
to have resolutions for puzzles which arise in
such a context for $\alpha$-vacua.  As a result, we will be brief in
discounting such problems in the following.

\subsection{Causality}

Na\"{\i}vely, because an exterior
point $x$ and its antipodal point $x_A$ are spacelike
separated, no causality problems arise from correlating the two
points.  More precisely, for de Sitter spaces, there are na\"{\i}vely
no causality problems because the lightcones emanating from $x$ and $x_A$
do not intersect.
However, it has been noted~\cite{el} that
this does not take backreaction into account.  Backreaction acts to
increase the de Sitter horizon and therefore makes de Sitter
``taller''.\cite{mw}  Then the lightcones emanating from $x$ and $x_A$
do intersect in the future, and this leads to causal problems for
$\alpha$-vacua.

\FIGURE[t]{
\includegraphics[width=3in,height=3.25in,clip=true,%
              keepaspectratio=true]{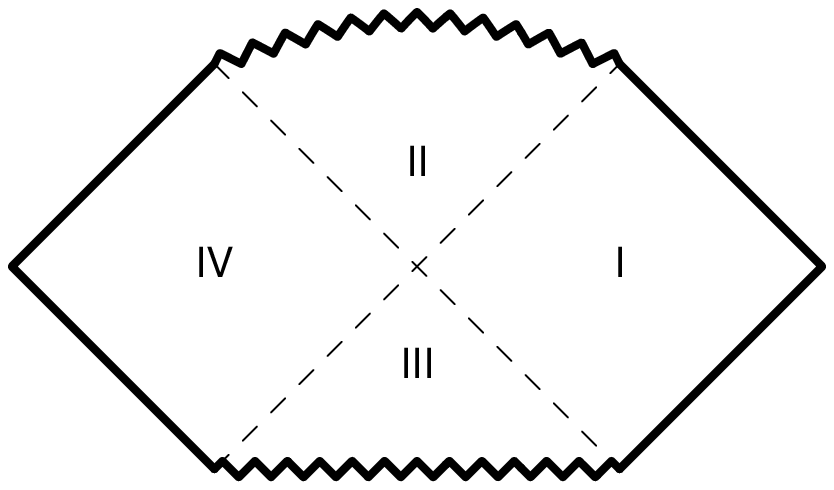}
\caption{Perturbations of Schwarzschild black holes increase the area
of the horizon.  On the Penrose diagram, this translates to the
depicted  deformation of the future singularity.
This particular diagram is the ``eternal Vaidya'' black hole
that is obtained by taking an eternal Schwarzschild black hole of mass
$M$ and
collapsing a spherical shell of radially directed null matter of additional
energy $M$, just to the future of the past horizon $v=-\infty$.%
\label{fig:pert}}
}

For black holes spacetimes, however, the backreaction will increase
the black hole horizon, thus increasing the amount of the space behind
the black hole horizons.  That is, the future singularity in
Fig.~\ref{fig:PD} arcs upwards (see Fig.~\ref{fig:pert}).
Thus, the causality problems that appear due to intersection of the
lightcones emanating from $x$ in region I and $x_A$ in region IV only
occur inside the black hole horizon.  Since we already do not understand
physics inside the horizon, this does not bother us.

\subsection{Poles}

In field theory, there are
singularities in tree-level three-point functions, $\vev{\phi(x_1)
  \phi(x_2) \phi(x_3)}$.  Namely, since the two-point functions are
singular for null separations (\cf\ Eq.~\eqref{HHG}),
and since the three-point function involves an integral,
\begin{equation}
\vev{\phi(x_1) \phi(x_2) \phi(x_3)} \sim \int d y G^F(x_1,y)
  G^F(x_2,y) G^F(x_3,y),
\end{equation}
there are poles when the integrated interaction point $y$ is on
the lightcone of one of the external points.
Of course, these poles are well-understood in terms of
physical propagating particles.
However, for $\alpha$-vacua, such poles also appear when $y$ is
on the lightcone of the antipodal of one of the external points.\cite{el}
This appears to be
unphysically acausal---in principle, it allows an observer to probe a
causally disconnected region, namely that of the antipodal points, if
all the points $x_1,x_2,x_3$ are causally connected.
Moreover, if one of the points happens to also lie on the lightcone of
another's the antipodal point, then for $y$ also on that lightcone,
there will be coincident poles leading to divergence of the
tree-level diagram.
These issues lead to questions about the sensibleness of
$\alpha$-vacua.

For black hole spacetimes, however, one again sees
from Fig.~\ref{fig:PD} that this problem only arises if at least one
of $x_1,x_2,x_3$ is inside the black hole horizon, as this is the only
place where the causal future of the exterior regions intersects.
Since we already
do not understand correlation functions between fields inside and
outside the horizon, this does not bother us.

\subsection{Thermality} \label{sec:AdStherm}

The nonthermality of Eq.~\eqref{Prat}
appears to be a disaster~\cite{kklss}.  Assuming a
steady state system, transitions from the $i$th state to the $j$th
state should be offset by transitions from the $j$th state to the
$i$th.  This is the statement of detailed balance:
\begin{equation} \label{db}
\rho(E_i) P(E_i \rightarrow E_j) = \rho(E_j) P(E_j\rightarrow E_i)
\Leftrightarrow 
\frac{P(E_i \rightarrow E_j)}{P(E_j \rightarrow E_i)} 
= \frac{\rho(E_j)}{\rho(E_i)}.
\end{equation}
For
the $\alpha=0$ vacuum, we conclude $\rho(E_i) \propto e^{-\beta E_i}$,
where $\beta$ is the inverse temperature, in appropriate units.  This
is just the Boltzmann distribution.

Indeed, as the authors of~\cite{kklss} have reminded us, detailed
balance~\eqref{db} implies the Boltzmann distribution.  Setting
\begin{equation}
R_{ij} \equiv 
  \frac{P(E_i \rightarrow E_j)}{P(E_j \rightarrow E_i)}
   \equiv \varphi(E_i-E_j),
\end{equation}
detailed balance implies that
\begin{equation}
R_{ij} R_{jk} = R_{ik},
\end{equation}
so if $\delta E\equiv E_j-E_i=E_k-E_j$ then
\begin{equation}
\varphi(\delta E)^2 = \varphi(2\delta E),
\end{equation}
which implies $\varphi(\delta E) = e^{\varphi'(0) \delta E}$.  That is
just a Boltzmann distribution, upon identifying $\beta=-\varphi'(0)$.

However, there are loopholes.  For one, we know that the Boltzmann distribution
is the classical distribution.  For Fermi-Dirac statistics, for example,
the transition from the $i$th state to the $j$th state cannnot occur
of the $j$th state is already occupied.  That is, Eq.~\eqref{db} is
replaced by
\begin{equation}
\rho(E_i) \left[1-\rho(E_j)\right] P(E_i \rightarrow E_j) 
= \rho(E_j) \left[1-\rho(E_i)\right] P(E_j\rightarrow E_i)
\end{equation}
or
\begin{equation}
\frac{P(E_i \rightarrow E_j)}{P(E_j \rightarrow E_i)} 
= \frac{\rho(E_j) \left[1-\rho(E_i)\right]}{\rho(E_i)
     \left[1-\rho(E_i)\right]}.
\end{equation}
Indeed, applying this to the $\alpha=0$ result~\eqref{P0rat} gives
\begin{equation}
\rho(E_i) \propto \frac{e^{-\beta E_i}}{1+e^{-\beta E_i}},
\end{equation}
which is the familiar Fermi-Dirac distribution.  The Bose distribution
is similarly obtained by realizing that the transition for bosons is
enhanced if the excited state is already occupied.\cite{feynman}

Nevertheless, it is clear that for a general $\alpha$-vacuum, the
result~\eqref{Prat} will not correspond to a familiar distribution.
For one, the expression does not factorize into a product of functions
of $E_i$ and $E_j$!  This is also not a disaster, however.
Most conservatively, we need not presume that the distribution in
question is that of an equilibrium system.  It is well-known (see
\eg~\cite{ruelle}) that the late time limit of any nonequilibrium
distribution function is a time-independent nonequilibrium steady-state
distribution function.  This need not be a familiar distribution; it
need only solve the Fokker-Planck equation (or the appropriate quantum
generalization thereof) which, being a second-order
differential equation, has a time-independent solution that is not the
Boltzmann distribution.  That is apparently the sitution here.
A concrete example of such a situation appears in astrophysics~\cite{me}.

An additional motivation for realizing that the result~\eqref{Prat} corresponds
to a nonequilibrium steady-state system is
the following. Consider the instanton that is the Euclidean black hole.
Na\"{\i}vely, its temperature is
determined by a length scale, namely the radius of the
instanton.  For the $\alpha$-vacua,
modes are explicitly correlated between the two ``sides'' of the space time,
permitting a ``short cut'' across the instanton.
Presumably, this affects the thermality.

However, we cannot resist mentioning that one can
obtain a familiar thermal distribution by replacing
the detailed balance relation~\eqref{db}
with the
peculiar expression%
\footnote{This is clearly not unique.}
\begin{multline} \label{weirdstats}
\abs{\cosh \alpha \sqrt{\rho(E_i)\left[1-\sigma \rho(E_j)\right]}
   + e^{i \gamma} \sinh \alpha \sqrt{\rho(E_j)\left[1-\sigma\rho(E_i)\right]}}^2
P(E_i\rightarrow E_j)\\
= \abs{\cosh \alpha \sqrt{\rho(E_j)\left[1-\sigma\rho(E_i)\right]}
   + e^{i \gamma} \sinh \alpha \sqrt{\rho(E_i)\left[1-\sigma\rho(E_j)\right]}}^2
P(E_j\rightarrow E_i),
\end{multline}
which implies the distribution
\begin{equation}
\rho(E_i) \propto \frac{1}{e^{\beta E_i}-\sigma}.
\end{equation}
This is a Boltzmann distribution for $\sigma=0$, a Bose-Einstein
distribution for $\sigma=1$ and a Fermi-Dirac distribution for
$\sigma=-1$, all at temperature $\beta$.
Eq.~\eqref{weirdstats} would imply that the meaning of detailed balance---\ie\ the quantum 
statistics---is vacuum dependent, as for anyons.  However, there is still
at least one puzzle which follows from this idea---namely, 
for $\alpha \neq 0$ it would be
possible to excite the detector from state $\ket{i}$ to $\ket{j}$ even if
the initial population of the $\ket{i}$'th state was empty provided $\ket{j}$
was sufficiently populated---and though that puzzle may have a resolution
in terms of an interpretation of a transition from energy $E_j$ to $E_i$
at the antipodal point, we will not advocate this interpretation as we prefer
the nonequilibrium statistical mechanical interpretation of~\eqref{Prat}.

\subsection{Pinch Singularities} \label{sec:pinch}

It has been noted~\cite{el}---see~\cite{bfh} for a more general phrasing
of this result---that quantum field theory loops become
ill-defined for $\alpha$-vacua because of the emergence of pinch
singularities.
For example, consider the one-loop correction to the propagator
\begin{equation} \label{oneloop}
\begin{split}
\begin{aligned}
\text{
\setlength{\unitlength}{.01in} 
\begin{picture}(150,100)(-75,-50)  
\put(0,0){\circle{50}} 
\put(-25,0){\line(-1,0){50}}
\put(25,0){\line(1,0){50}}
\put(-25,-10){\llap{$x$}}
\put(25,-10){$y$}
\end{picture}
}\end{aligned}
&\sim \int dx \int dy \, G^F_{\alpha\gamma}(x,y) G^F_{\alpha\gamma}(y,x)\\
&\sim \ldots + 
\int dx \int dy \, \cosh^2 \alpha\, \sinh^2\alpha\, G^F_0(x,y) G_0^F(y,x)^*
   + \ldots.
\end{split}
\end{equation}
The term we have shown involves both $i\epsilon$ prescriptions, and
thus it is impossible to deform the contour of integration so as to
avoid the singularities.  More precisely, the residue of the pole from
any one factor is singular due to the pole in the other term.\cite{el}

For this reason, many authors~\cite{el2,gl,gl2,ch}
have suggested modifying the time ordering prescription for
$\alpha$-vacua,
in order to remove the pinch singularities.
But presumably string quantization works in spacetimes
with nonpositive cosmological constant, and it is well-known that
string loops are nonsingular.  Thus, we anticipate that for the black
hole spacetimes considered here, the pinch singularities can be
evaded without altering the time ordering prescription.

That this
anticipation is not ridiculous is seen by considering strings in
Rindler space.%
\footnote{\skipthis{Details will appear elsewhere.  }Past work on strings in
Rindler space includes~\cite{vs,ls} but we have not seen the
approach given here, which resembles some calculations of~\cite{lms},
included in the Rindler literature.}
Rindler space is just Minkowski space after a coordinate transformation, and
string $n$-point amplitudes
are just integrated worldsheet correlation functions,
\begin{equation}
\vev{{\mathcal V}_1 \cdots {\mathcal V}_n}.
\end{equation}
For tachyons in ordinary Minkowski space, the vertex operators are
just the Minkowski modes ${\mathcal V}_j = e^{i k_j \cdot X}$; in
Rindler space, they are the Rindler modes.  The latter, however, are
obtainable as a Bogliubov transformation of the former; that is,
string amplitudes in Rindler space can be obtained as
an appropriate integral (over $k_1,\dotsc, k_n$) of the Minkowski string
amplitudes.
This yields a finite result.  One can define
``$\alpha$-vacua'' of Rindler space using $PT$ as the ``antipodal map'',
though they are not as well-defined because
$PT$ has a fixed point at the origin of the Minkowski spacetime.
Although loop amplitudes in these ``$\alpha$-vacua'' would
na\"{\i}vely be expected to suffer pinch singularities due to the
reversal of Rindler time, in fact the string loop amplitudes are just
linear combinations of the ordinary, and finite, Rindler space
amplitudes, and so are finite.  We conclude that strings do not suffer
from pinch singularities.  The still-skeptical reader can find
details in Appendix~\ref{sec:rindler}.

One might still ask%
\footnote{J.M thanks the referee for bringing this omission in our
original argument to our attention.}
how it is that the field theory pinch singularities are recovered from
the finite string theory amplitudes.  For concreteness, we will consider
the open string in our Rindler toy model.
The open string one-loop annulus diagram
(\eg\ Fig.~\ref{fig:open2})
is conformally equivalent to a
cylinder, and includes an integral over the
modulus $t$, the ratio between the radius and the length of the cylinder.
The field theory limit is obtained in the limit $t\rightarrow \infty$, in
which the open string loop becomes infinitely short.  The annulus diagram
also includes integrals over the positions of the vertex operators which
live on the boundary; these
range over the circumference of the cylinder, $2\pi t$.

\FIGURE[t]{
\includegraphics[width=5in]{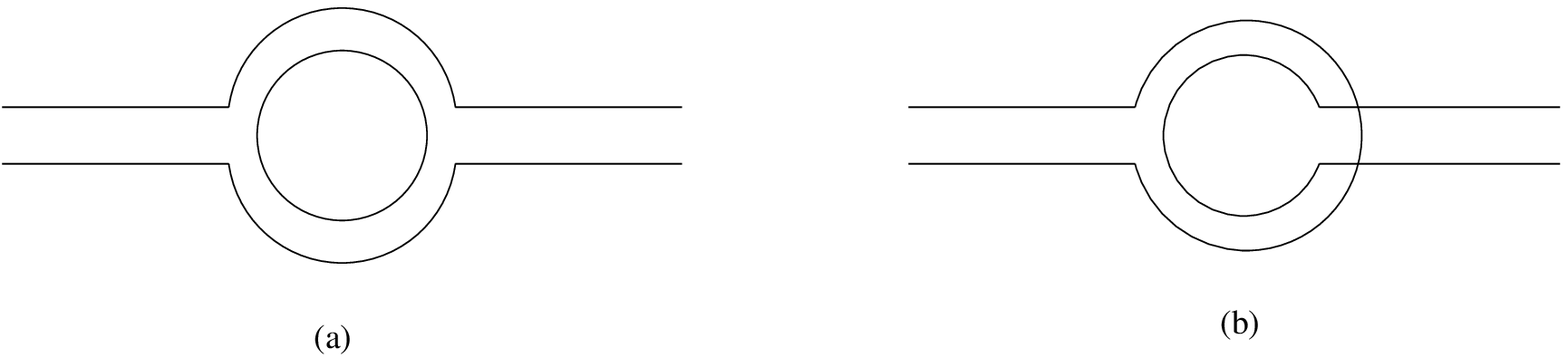}
\caption{The (a) planar and (b) nonplanar 1-loop open string 2-point
diagrams.  As {\em position space field theory\/} diagrams (take
widths to zero) the two internal propagators connecting the two
vertices, when expanded in an $\alpha$-vacuum, have cross-terms that
give the pinch singularities.  However, pulling out the right-hand
external state for the nonplanar diagram (b) gives a single closed
string propagator, and therefore no pinch singularity.
So any pinch singularity must come from the opposite limit,
which is the field theory limit, $t\rightarrow\infty$.\label{fig:open2}}
}

By considering open string factorization of the annulus diagram,
(this argument follows~\cite{lm} most closely and is reproduced
in Appendix~\ref{sec:osfac},
although this result dates back to at least~\cite{jpfac})---namely,
that the one-loop
annulus amplitude can be reconstructed from the tree-level amplitude on the
plane by insertion of two open string vertices in the plane, as depicted
in Fig.~\ref{fig:openfac}---one can recover the propagator of the
intermediate particle.
The additional propagator factor which leads to the pinch singularity
is obtained from the integral over
the vertex operator positions as well.  However, the pinch singularity 
is only obtained because
in the $t\rightarrow\infty$ limit, the vertex operator position extends, in
the coordinates of Appendix~\ref{sec:osfac}, to $y=\pm 1$.  For finite $t$, the
region of integration for the vertex operator position integral is shrunk
to $y=\pm \tanh \frac{\pi t}{2}$,
and no singularity appears.  In this way, string theory regularizes
the field theory pinch singularity.

\FIGURE[t]{
\includegraphics[width=5in]{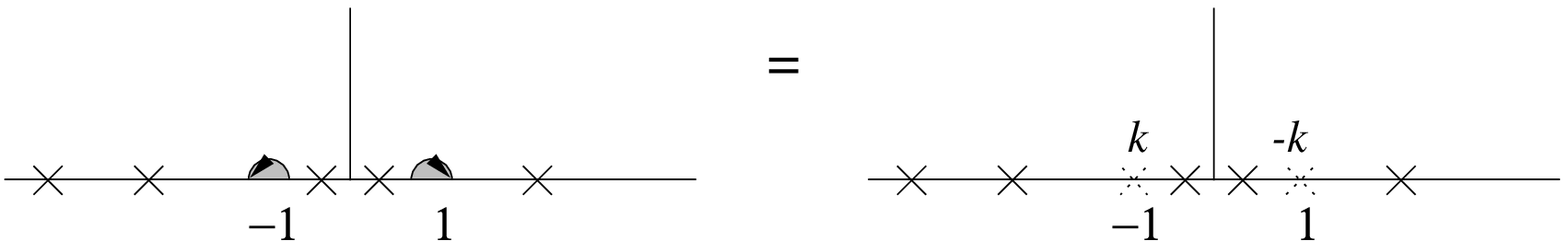}
\caption{Open string factorization
of the annulus occurs as the modular parameter of the annulus
$t\rightarrow\infty$, corresponding to an infinitely short cylinder.
For a general value of the modulus, the annulus diagram is equivalent to
a sum of tree-level diagrams in which one inserts vertex operators with
equal and opposite momentum and cuts out, and identifies, a region around
the inserted vertex operators.  The sum is over the momentum and identity
of the inserted vertex operators; this corresponds to the loop exchange
of a virtual string.
This figure was
stolen from~\cite{lm}.\label{fig:openfac}}
}

On this note, it is intriguing to recall
(see \eg~\cite{nopinch1,nopinch2,nopinch3})
that na\"{\i}vely expected pinch singularities in nonequilibrium
field theory fail to appear after careful calculation if the
interaction that leads to the pinch singularity is turned on for only
a finite time.  Since the integration over the location
of the string theory vertex operator is an integration over 
a worldsheet time coordinate, the finiteness of the modulus $t$, away
from the field theory limit, corresponds to finiteness of the worldsheet time
and so could be precisely this mechanism.  We hope that further work
will shed light on this.

\subsection{Uniqueness of Black Hole Vacua}
It is known~\cite{pc}
that of the usually considered black hole vacua, only the
Hartle-Hawking vacuum enjoys the property
that the renormalized stress tensor does not diverge on either
the past or future black hole horizon.
Does this not rule out $\alpha$-vacua?\footnote{J.M. thanks Rob Myers
for asking
this interesting question at the Andrew Chamblin Memorial Symposium
at the University of Louisville in Kentucky, USA.}

Actually, the stress tensor does not diverge on the horizon in any 
$\alpha$-vacuum constructed from the Hartle-Hawking vacuum.%
\footnote{Similarly, the stress tensor diverges on the past but not
the future horizon in any
$\alpha$-vacuum constructed from the Unruh vacuum.}
The point is that
the renormalization of the stress tensor is related to
$G^F_{\alpha\gamma}(x,x)$, where
$x$ is evaluated on the black hole horizon.  
Using~\eqref{GFalpha},
this will be finite if the Hartle-Hawking Green functions
$G^F_0(x,x)$, $G^{(1)}_0(x,x_A)$ and $D_0(x,x_A)$ are finite.
The work of Candelas~\cite{pc} demonstrated that $G^F_0(x,x)$ is finite
on the horizon.  Moreover,
for $x$ on the black (white) hole horizon, $x_A$ is on the white (black) hole
horizon; on the Penrose diagram,
the points appear to be null-like related, but because the antipodal map
includes the antipodal map on the sphere, $x$ and $x_A$ are
therefore actually spacelike related for $x$ on the horizon.  Thus
$D_0(x,x_A)=0$, and $G^{(1)}_0(x,x_A)=G^F_0(x,x_A)$ is finite.
So the renormalized stress tensor is finite on the horizon for all
$\alpha$-vacua constructed from the Hartle-Hawking vacuum.

This does not contradict Candelas' result, as he 
only considered the Boulware, Hartle-Hawking and Unruh
vacua.

\section{CFT $\alpha$-States} \label{sec:CFTalpha}

Given that Schwarzschild-AdS has two boundaries and
$\alpha$-vacua, there should be a one-parameter set of CFT states.
We provide a proposal in this section.

First, though, we should point out that this proposal differs
substantially from that for dS/CFT~\cite{bms}.  For dS/CFT, the
authors noted that, unlike AdS/CFT%
\footnote{The exception is the special range,
$-\frac{d^2}{4} < m^2 < -\frac{d^2}{4}+1$,
of AdS$_{d+1}$ tachyon masses
for which the two CFT operators, $\op_\pm$,
correspond to two inequivalent bulk
quantizations of the scalar field.\cite{kw}
In particular, it is argued in~\cite{kw} that the two
inequivalent quantizations of AdS correspond to two different CFTs,
one with only $\op_+$ and one with only $\op_-$.
So even in this special range, the putatively
marginal $\op_+ \op_-$ does not exist in AdS.
And even if this operator did exist in this case,
it would be unsatisfying to have a CFT
$\alpha$-state prescription that only held in this limited range of
AdS masses.%
\label{ft:AdSO+O-}}%
, each dS mode is associated with two CFT operators, the product of
which is marginal.  Thus, the dS $\alpha$-vacua can be associated
with the CFT deformed by this marginal operator.  For AdS, however,
because only one of
the two candidate CFT operators actually exists$^{\text{\ref{ft:AdSO+O-}}}$,
this prescription does not apply.

So let us recall the prescription~\cite{juan,fhks} for the $\alpha=0$
vacuum.
The two
boundary CFTs
give a product Hilbert space $\ket{}_{1}\ket{}_{2}$.  Because time runs in
opposite directions on the two boundaries, one has, just as in
real-time thermal field theory~(for a review see \cite{thermal}),
\hbox{$H=H_1\otimes \one - \one \otimes H_2$}.
Thus the state
\begin{equation} \label{psi}
\ket{\psi}_0 = \frac{1}{\sqrt{Z}} \sum_i e^{-\beta E_i/2} \ket{i}_1 \ket{i}_2,
\end{equation}
written in terms of a complete set of energy eigenstates of each CFT,
has vanishing energy.  It is unit normalized using the partition
function $Z$.  It cannot be overemphasized that it is a {\em pure state\/}.

The two identical boundary CFTs are related by an antilinear involutive
map~\cite{thermal}, with the result that
\begin{subequations} \label{op2to1}
\begin{align}
e^{\beta H/2} \left[\one \otimes \op(t,\Omega)\right] \ket{\psi}_0
&= \left[\op^\dagger(t,\Omega) \otimes \one\right] \ket{\psi}_0, \\
\intertext{or equivalently, and somewhat more schematically,}
\op_2(t,\Omega) \ket{\psi}_0 &=
\op_1^\dagger(t-i\tfrac{\beta}{2}) \ket{\psi}_0,
\end{align}
\end{subequations}%
Here, 
$\op_1^\dagger(t-i\tfrac{\beta}{2})=\left[e^{H_1 \beta/2} \op_1(t)
e^{-H_1\beta/2}\right]^\dagger$ is the Hermitian conjugate of
the operator evaluated at time \hbox{$t-i\tfrac{\beta}{2}$}
and so is the Hermitian conjugate
of the operator, evaluated at time 
\hbox{$t+i\frac{\beta}{2}$}%
.
As a result, one has the relation, for operators $A,B$ in each field
theory, (\cf~\cite{fhks})
\begin{equation}
\bra{\psi}_0 A_1(t) B_2(t') \ket{\psi}_0
= \bra{\psi}_0 A_1(t) B_1^\dagger(t'-i\tfrac{\beta}{2}) \ket{\psi}_0.
\end{equation}

However, observers on one boundary or the other can only see their own
boundary.  Such an observer, in
CFT$_1$ for definiteness, will not see boundary 2, and so its
observations will involve a trace over the CFT$_2$ Hilbert space.
That is, an observer in CFT$_1$ sees the density matrix
\begin{equation}
\rho_1 = \Tr_2 \ket{\psi} \bra{\psi} = \frac{1}{Z} \sum_i e^{-\beta
  E_i} \ket{i}_1 \; {_1}\bra{i}.
\end{equation}
This is precisely the thermal density matrix.

The Schwarzschild-AdS $\alpha$-vacua are constructed by a
Bogoliubov transformation that correlates antipodal points.  In the
CFT, then, this suggests an analogous Bogoliubov transformation.  We
will make this precise for a single one-dimensional, harmonic
oscillator on each ``boundary'';
the generalization to actual dual CFTs should be obvious by
replacing the harmonic oscillator creation and annihilation operators
by operators which create and annihilate the state $\ket{i}$ on each boundary.

We have two copies of the harmonic oscillator,
\begin{equation}
\com{a_1}{a_1^\dagger} = 1 = \com{a_2}{a_2^\dagger}.
\end{equation}
The pure state above can be written as
\begin{equation}
\ket{\psi}_0 = \sqrt{1-e^{-\beta}} 
\exp\left[e^{-\frac{\beta}{2}} a_1^\dagger a_2^\dagger\right]
 \ket{0}_1 \ket{0}_2,
\end{equation}
after setting the harmonic oscillator frequency 
$\omega=1$ (otherwise $\beta\rightarrow
\beta\omega$) and using
\begin{equation}
Z = \half \csch \frac{\beta}{2}.
\end{equation}

We can now perform the Bogoliubov transformation,
\begin{align} \label{Bops}
b_1^\dagger &= \cosh \alpha\, a_1^\dagger - e^{i\gamma} \sinh\alpha\, a_2,
&
b_2^\dagger &= \cosh\alpha\, a_2^\dagger - e^{i\gamma} \sinh\alpha\, a_1.
\end{align}
Note that this preserves the Hamiltonian
\begin{equation}
H = a_1^\dagger a_1 - a_2^\dagger a_2 = b_1^\dagger b_1 - b_2^\dagger b_2,
\end{equation}
in accord with the statement that $AdS$ $\alpha$-vacua preserve the
symmetries preserved by the standard (Hartle-Hawking) vacuum.
The new operators define new vacua via
\begin{equation}
b_1 \nvo = 0 = b_2 \nvt.
\end{equation}
Note that although the ``CFT'' is still a product CFT,
it is no longer manifestly
a product of CFTs on each boundary.

We now use the Bogoliubov transformed operators to build a new pure state,
$\ket{\psi}_{\alpha\gamma}$, namely
\begin{equation} \label{apsi}
\ket{\psi}_{\alpha\gamma} = \sqrt{1-e^{-\beta}} \exp\left[e^{-\frac{\beta}{2}}
  b_1^\dagger b_2^\dagger\right] \nvo \nvt.
\end{equation}
\iftoomuchdetail
\begin{detail}%
Let us check that this is indeed $\alpha$-dependent.  Standard
arguments \cite{bms,el2} and references therein, allow us to write
\begin{equation} \label{detailsqueeze}
\nvo \nvt = \sech \alpha
   e^{e^{-i \gamma}\tanh\alpha \, a_1^\dagger a_2^\dagger} \ket{0}_1 \ket{0}_2.
\end{equation}
This is independent of $\beta$, and so it is almost obvious (and
becomes obvious upon considering $\beta\rightarrow \infty$) that
multiplying by the
operator in Eq.~\eqref{apsi} will not result in Eq.~\eqref{psi}.
So we have indeed found a candidate $\alpha$-state in the CFT.
Like $\ket{\psi}_0$, the state $\ket{\psi}_{\alpha\gamma}$ has vanishing
(total) energy.

\end{detail}%
\fi
To what density matrix does this state correspond?  Again, we consider
an observer on the first boundary.  Such an observer does not observe
the second boundary, and so we should again trace over CFT$_2$.
This is {\em not\/} the same as tracing over the
$(b_2^\dagger,b_2)$ Hilbert space!  That Hilbert space partly exists
on CFT$_1$.  That is,
\begin{equation}
\Tr_2 \neq \widetilde{\Tr}_{{2}}.
\end{equation}

In Appendix~\ref{sec:getrho} we show that
\begin{equation} \label{gotpsi}
\ket{\psi}_{\alpha\gamma}
= \frac{\sech \alpha \,\sqrt{1-e^{-\beta}}}{
      1+e^{-\frac{\beta}{2} + i\gamma}\tanh \alpha}
  \exp \left[ \frac{e^{-\frac{\beta}{2}} + e^{-i\gamma} \tanh \alpha}{
       1 + e^{-\frac{\beta}{2} +i\gamma} \tanh\alpha} 
       a_1^\dagger a_2^\dagger \right]
\ket{0}_1 \ket{0}_2.
\end{equation}
We should note, following~\cite{el2}, that this makes the state associated
with $\alpha$-vacua appear to be in the same Hilbert space as the
state associated with the Hartle-Hawking vacuum.  However, the CFT
associated with an AdS black hole has an infinite number of
oscillators; then the inner product
${_{\alpha\gamma}}\braket{\psi}{\psi}_0 \sim \sech^\infty \alpha = 0$.
This argument is easily generalized to any state built up from
$\ket{\psi}_{\alpha\gamma}$ and so
states for different values of $\alpha,\gamma$
are in different Hilbert spaces.

Eq.~\eqref{gotpsi} yields the density matrix
\begin{equation} \label{gotrho}
\begin{split}
\rho_{\alpha\gamma} &=
\Tr_2 \ket{\psi}_{\alpha\gamma} \; {_{\alpha\gamma}}\bra{\psi}
\\ &= (1-e^{-\beta}) 
\frac{\sech^2 \alpha}{
   \abs{1+e^{-\frac{\beta}{2}+i\gamma} \tanh \alpha}^2}
\sum_{n=0}^{\infty} e^{-\beta n}
\abs{\frac{\cosh \alpha + e^{\frac{\beta}{2}+i\gamma} \sinh \alpha}{
  \cosh\alpha + e^{-\frac{\beta}{2} + i\gamma} \sinh \alpha}}^{2n}
  \ket{n}_1 \; {_1}\bra{n}
\end{split}
\end{equation}
Note that at $\alpha=0$, this reduces to the standard thermal answer.
Also, this shows that the density matrix
is $\alpha$-dependent, and thus the entropy 
$S_{\alpha\gamma} = -\Tr \rho_{\alpha\gamma} \ln \rho_{\alpha\gamma}$
is $\alpha$-dependent.

Explicitly,
\begin{equation} \label{Sa}
S_{\alpha\gamma} = - \ln \left[ \frac{(1-e^{-\beta}) \sech^2\alpha}{
   \abs{1 + e^{-\frac{\beta}{2}-i\gamma}\tanh\alpha}^2} \right]
- \frac{\abs{e^{-\frac{\beta}{2}} +e^{i\gamma} \tanh\alpha}^2 \cosh^2\alpha}{
      1-e^{-\beta}}
  \ln \frac{\abs{e^{-\frac{\beta}{2}} + e^{i\gamma} \tanh\alpha}^2}{
       \abs{1+e^{-\frac{\beta}{2}-i\gamma} \tanh \alpha}^2}.
\end{equation}
It is interesting to see how $\alpha$ affects the high and low
temperature entropy.  This is shown in Fig.~\ref{fig:entropy}.

\FIGURE[t]{
\begin{tabular}{cc}
\subfig{\includegraphics[height=2in,clip=true,%
              keepaspectratio=true]{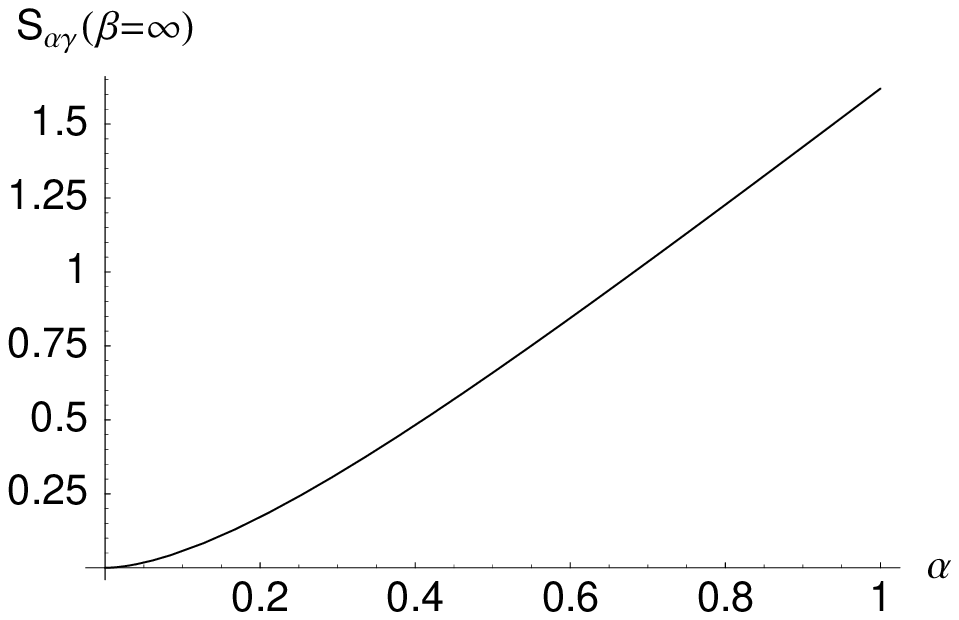}}%
    \label{fig:entropy:LowT} &
\subfig{\includegraphics[height=2in,clip=true,%
              keepaspectratio=true]{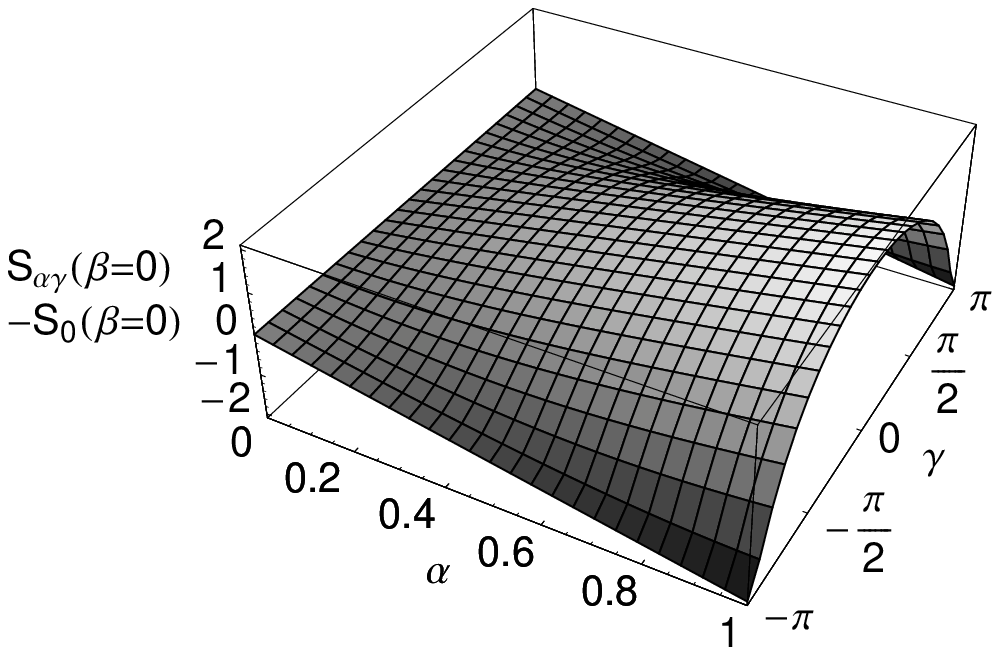}}%
    \label{fig:entropy:HighT}
\end{tabular}
\caption{The $\alpha$ dependence of the entropy at
extreme~\ref{fig:entropy:LowT}
low temperature and~\ref{fig:entropy:HighT} high temperature.
\label{fig:entropy}
}}

\section{AdS/CFT} \label{sec:AdSCFT}

\subsection{Populations} \label{sec:AdSCFTbalance}

Let us compare Eq.~\eqref{gotrho} and~\eqref{Prat}.
To do this, we should make the usual identification between the
harmonic oscillator occupation number, and the number of particles in
a given state.  Thus, [upon restoring the frequency to
  Eq.~\eqref{gotrho} via $\beta\rightarrow \beta E_i$]
the ratio of probabilities between having $n$ particles in the
$i^\text{th}$ state and $n-1$ particles in that CFT state, is
\begin{equation} \label{CFTrat}
\frac{\rho(n)}{\rho(n-1)} 
= e^{-\beta E_i}
\abs{\frac{\cosh \alpha + e^{\frac{\beta E_i}{2}+i\gamma} \sinh \alpha}{
  \cosh\alpha + e^{-\frac{\beta E_i}{2} + i\gamma} \sinh \alpha}}^{2},
\end{equation}
which exactly matches the ratio of transition rates of the bulk
detector~\eqref{Prat}!  

\subsection{Propagators}

Now let us make the prescription of $\S$\ref{sec:CFTalpha}
somewhat more precise.
In~\cite{bkl,bklt} it is argued that the mode expansion of the
bulk field $\phi$,
\begin{equation} \label{phimodes2}
\phi(x) = \sum_{\kappa,n,s^{(n)}} \left[
   a^{\vphantom{\dagger}}_{\kappa,n,s^{(n)}} 
  \phi^{\vphantom{*}}_{\kappa,n,s^{(n)}}(x)
 + a^\dagger_{\kappa,n,s^{(n)}} \phi^*_{\kappa,n,s^{(n)}}(x) \right],
\end{equation}
implies the mode expansion of the dual operator
\begin{equation} \label{bdymodes}
\op(t,\Omega) = \sum_{\kappa,n,s^{(n)}} \left[
  b^{\vphantom{\dagger}}_{\kappa,n,s^{(n)}}
    \tilde{\phi}^{\vphantom{*}}_{\kappa,n,s^{(n)}}(t,\Omega)
+ b^{\dagger}_{\kappa,n,s^{(n)}}
    \tilde{\phi}^{*}_{\kappa,n,s^{(n)}}(t,\Omega)
\right],
\end{equation}
where the mode $\tilde{\phi}_{\kappa,n,s^{(n)}}$ is related to the
boundary value of $\phi_{\kappa,n,s^{(n)}}$ and $\Omega$ parametrizes
the boundary $S^{d-1}$.
With two boundaries, this is more subtle.  We would like to write, on
each boundary,
\begin{equation} \label{2bdymodes}
\begin{aligned}
\op_1(t,\Omega) &= \sum_{\kappa,n,s^{(n)}} \left[
  b^{\vphantom{\dagger}}_{1\kappa,n,s^{(n)}}
    \tilde{\phi}^{\vphantom{*}}_{\kappa,n,s^{(n)}}(t,\Omega)
+ b^{\dagger}_{1\kappa,n,s^{(n)}}
    \tilde{\phi}^{*}_{\kappa,n,s^{(n)}}(t,\Omega)
\right], \\
\op_2(t,\Omega) &= \sum_{\kappa,n,s^{(n)}} \left[
  b^{\vphantom{\dagger}}_{2\kappa,n,s^{(n)}}
    \tilde{\phi}^{\vphantom{*}}_{\kappa,n,s^{(n)}}(t-i\tfrac{\beta}{2},\Omega)
+ b^{\dagger}_{2\kappa,n,s^{(n)}}
    \tilde{\phi}^{*}_{\kappa,n,s^{(n)}}(t-i\tfrac{\beta}{2},\Omega)
\right],
\end{aligned}
\end{equation}
but the bulk modes, restricted to each boundary, form an overcomplete
set of functions on each boundary.  Basically, the bulk modes come in
pairs, one linear combination of which vanishes on one boundary and
another of which vanishes on the other boundary.
However, we can write Eq.~\eqref{2bdymodes} assuming that we have
restricted the sums to a complete set of functions, and
such that, just as for the bulk theory, we have
\begin{equation}
\tilde{\phi}_{\kappa,n,s^{(n)}}^*(t,\Omega)
 = \tilde{\phi}_{\kappa,n,s^{(n)}}(t-i\tfrac{\beta}{2},\Omega_A),
\end{equation}
where $\Omega_A$ is the antipodal point on the sphere to $\Omega$, and
$t-i\frac{\beta}{2}$ is the corresponding time on the other boundary.
Moreover,~\cite{bdhm} (see also
\cite{bkl,bklt,br,kos})
argue that the quantum numbers that we have incorporated into $\kappa$
are essentially the energy (after all, there is a time-like Killing
vector) and that the relationship Eqs.~\eqref{phimodes2}
and~\eqref{bdymodes} implies that the two-point function
$\vev{\op(t,\Omega) \op(t',\Omega')}$ is given by the bulk
propagator.  We would now like to check that this holds as well after
the Bogoliubov transformation.

First, however, we should note that since the operators in the mode
expansion of the composite operator~\eqref{bdymodes} have well-defined
energies, that the compositeness of the operator to which the creation
and annihilation operators are defined is not incompatible with our previous
discussions in~$\S$\ref{sec:CFTalpha},~$\S$\ref{sec:AdSCFTbalance}.
Indeed, it will be sufficient to discuss a single set of oscillators,
so that
\begin{equation}
\begin{aligned}
\op_1 &= A_1(t,\Omega) a_1 + A_1(t,\Omega)^* a^\dagger_1 
\\ &= A_1 \left(\cosh \alpha \, b_1 
 + e^{i\gamma} \sinh \alpha \, b_2^\dagger \right)
+ A_1^* \left( \cosh\alpha \, b^\dagger_1
 +  e^{-i\gamma} \sinh \alpha\, b_2 \right),
\\
\op_2 &= A_2(t-i\tfrac{\beta}{2},\Omega) a_2 
  + A_2(t-i\tfrac{\beta}{2},\Omega)^* a^\dagger_2
\\ &= A_2 \left(\cosh\alpha \, b_2 
+ e^{i\gamma} \sinh \alpha \, b_1^\dagger \right)
+ A_2^* \left(\cosh\alpha \, b_2^\dagger 
   +  e^{-i\gamma} \sinh \alpha \, b_1 \right).
\end{aligned}
\end{equation}

Let us note that
\begin{equation}
\begin{aligned}
{_{\alpha\gamma}}\bra{\psi}b_1 b_1^\dagger \ket{\psi}_{\alpha\gamma}
&= \thalf e^{\tfrac{\beta}{2}} \csch \tfrac{\beta}{2}, &
{_{\alpha\gamma}}\bra{\psi}b_1^\dagger b_1 \ket{\psi}_{\alpha\gamma}
&= \thalf e^{-\tfrac{\beta}{2}} \csch \tfrac{\beta}{2}, &
{_{\alpha\gamma}}\bra{\psi}b_1 b_2 \ket{\psi}_{\alpha\gamma}
&= \thalf \csch \tfrac{\beta}{2},
\end{aligned}
\end{equation}
Therefore, for $t>t'$,
\begin{multline}
{_{\alpha\gamma}}\bra{\psi} \op_1(t,\Omega) \op_1(t',\Omega')
\ket{\psi}_{\alpha\gamma}
= \cosh^2\alpha \,
     {_0}\bra{\psi} \op_1(t,\Omega) \op_1(t',\Omega') \ket{\psi}_0
\\
+ \sinh^2\alpha \,
     {_0}\bra{\psi} \op_1(t,\Omega) \op_1(t',\Omega') \ket{\psi}_0^*
+ \cos\gamma \sinh 2\alpha \,
     {_0}\bra{\psi} \op_1(t,\Omega) \op_1(t'-i\tfrac{\beta}{2},\Omega'_A)
     \ket{\psi}_0.
\end{multline}
The form of the last term arises by identifying
$A_1(t,\Omega)=A_1(t-i\tfrac{\beta}{2},\Omega_A)^*$
and by noting that the correlation function between the point
$(t,\Omega)$ and the ``antipodal'' point $(t'-i\tfrac{\beta}{2},\Omega'_A)$
should be identified with
${_0}\bra{\psi}\op_1(t,\Omega)\op_2(t',\Omega')\ket{\psi}_0$; in particular,
since this corresponds to the bulk propagator between spacelike
points, it should be identified with the Hadamard function.
Thus, this agrees precisely with Eq.~\eqref{GFalpha}.  The last term
there (proportional to $\sin\gamma$) involves the commutator
function, which vanishes for spacelike separated points.

By analytic continuation then---in fact, precisely the analytic
continuation that allowed us to match the $\cos \gamma$ term---we
find that
\begin{equation}
{_{\alpha\gamma}}\bra{\psi} \op_1(t,\Omega) \op_2(t',\Omega')
\ket{\psi}_{\alpha\gamma}
\end{equation}
also matches the bulk propagator between the two boundaries.  However,
direct computation shows agreement only for $\gamma=0,\pi$.  (For
example, the coefficient of $\sinh^2 \alpha$ involves $e^{\pm 2 i
\gamma}$ instead of being $\gamma$-independent.)  We suspect that this
is related to the CPT  noninvariance of the $\gamma\neq 0,\pi$ vacua.

\subsection{Entropy}

The ``CFT calculation''~\eqref{Sa} now shows that the entropy of the
bulk is $\alpha$-dependent.  Thus, black hole entropy need not be $A/4$!
It would be interesting to understand this better.

In particular, this result for the black hole entropy assumes first of all
that an equilibrium calculation of the entropy holds for these nonequilibrium
(but steady state) $\alpha$-vacua.  This need not be true.

Additionally, na\"{\i}ve application of the equilibrium thermodynamic formula
\begin{equation} \label{thermo}
\frac{dS}{dE} = \beta,
\end{equation}
leads to the fairly nonsensical, energy-dependent temperature%
\footnote{But note that plugging this into the Boltzmann formula does in
fact reproduce Eq.~\eqref{Prat}!}
\begin{equation}
\beta_{\alpha\gamma}(E_i) = -\ln \abs{
\frac{e^{-\frac{\beta E_i}{2}} \cosh\alpha + e^{i\gamma} \sinh\alpha}{
  \cosh \alpha + e^{-\frac{\beta E_i}{2} + i \gamma}
  \sinh\alpha}}^2.
\end{equation}
Again, the obvious resolution is that the standard equilibrium formulas
do not apply in this nonequilibrium situation.
This leaves as an open question whether the black hole entropy is, or
is not, $A/4$ for all $\alpha$, but this is a question that can be addressed
using the dual CFT.

\acknowledgments

It is a pleasure to thank Sumit Das, Moshe Elitzur, Gary Gibbons, Finn
Larsen, Andreas Karch, Don Marolf, Matt Martin, Robert McNees, Emil
Mottola, Simon Ross, Al Shapere, Anastasia Volovich and Xinkai Wu for
useful conversations.  This work was supported in part by National
Science Foundation grant Nos PHY-0244811 and PHY-0555444 and by a
Department of Energy contract No. DE-FG01-00ER45832.  At The Ohio
State University, J.M. is supported in part by Department of Energy
contract No. DE-FG02-91ER-40690.

\appendix

\section{Derivation of the CFT Density Matrix} \label{sec:getrho}

In this appendix we derive the density matrix~\eqref{gotrho}.
Standard
arguments \cite{bms,el2} and references therein, allow us to write
\begin{equation} \label{squeeze}
\nvo \nvt = \sech \alpha \,
   e^{e^{-i \gamma}\tanh\alpha \, a_1^\dagger a_2^\dagger} \ket{0}_1 \ket{0}_2.
\end{equation}

Now we state the main formula we need,
which is a version of~\cite[(A5.17)]{optics}.
Suppose we have a set of operators $K_\pm, K_3$ obeying
\begin{align}
\com{K_3}{K_\pm} &= \pm K_\pm, & \com{K_+}{K_-}=-2K_3.
\end{align}
We will use
\begin{align}
K_+ &= a_1^\dagger a_2^\dagger, & K_- &= a_1 a_2, &
K_3 &= \half(a_1^\dagger a_1 + a_2 a_2^\dagger),
\end{align}
(note the ordering in $K_3$).  Then, for c-numbers
$\gamma_\pm$, $\gamma_3$ related by
\begin{equation}
\frac{1}{4} \gamma_3^2 = \gamma_+ \gamma_-,
\end{equation}
and setting,
\begin{align}
\Gamma_\pm &= \frac{\gamma_\pm}{1 - \frac{\gamma_3}{2}}, &
\Gamma_3 &= \frac{1}{\left(1-\frac{\gamma_3}{2}\right)^2},
\end{align}
one has
the identity,
\begin{equation} \label{opticsfmla}
\exp \left[ \gamma_+ K_+ - \gamma_3 K_3 + \gamma_- K_- \right]
= \exp \left[ \Gamma_+ K_+ \right]
  \exp \left[ (\log \Gamma_3) K_3 \right]
  \exp \left[ \Gamma_- K_- \right].
\end{equation}

We apply this first to
\begin{equation}
\exp \left[ e^{-\frac{\beta}{2}} b_1^\dagger b_2^\dagger \right]
= \exp \left[ e^{-\frac{\beta}{2}} \cosh^2\alpha \, K_+
  - 2 e^{-\frac{\beta}{2}} e^{i\gamma} \cosh\alpha\, \sinh\alpha\, K_3
  + e^{-\frac{\beta}{2}} e^{2i\gamma} \sinh^2\alpha\, K_- \right].
\end{equation}
so that,
\begin{equation}
\ket{\psi}_{\alpha\gamma} = \sech \alpha \sqrt{1-e^{-\beta}}
  e^{\Gamma_+ a^\dagger_1 a^\dagger_2} e^{(\log\Gamma_3) K_3} 
  e^{\Gamma_- a_1 a_2}
  e^{e^{-i \gamma} \tanh \alpha\, a_1^\dagger a_2^\dagger} 
\ket{0}_1 \ket{0}_2, \\
\end{equation}
We can now commute the right-most factor to the left, though this
requires another application of~\eqref{opticsfmla}.  This gives,
\begin{equation}
\ket{\psi}_{\alpha\gamma}
= \frac{\sech \alpha \,\sqrt{1-e^{-\beta}}}{
      1+e^{-\frac{\beta}{2} + i\gamma}\tanh \alpha}
  \exp \left[ \frac{e^{-\frac{\beta}{2}} + e^{-i\gamma} \tanh \alpha}{
       1 + e^{-\frac{\beta}{2} +i\gamma} \tanh\alpha} 
       a_1^\dagger a_2^\dagger \right]
\ket{0}_1 \ket{0}_2,
\end{equation}
which immediately implies Eq.~\eqref{gotrho}.

As a (trivial) check, one can see that setting $\alpha=0$ reproduces
the standard density matrix, and that in general $\Tr \rho_{\alpha\gamma} = 1$.

\section{Strings on Rindler Space} \label{sec:rindler}

In this section, we elaborate on the claims made in $\S$\ref{sec:pinch}
that strings do not suffer from pinch singularities in Rindler space.
Many of the field theory details we provide below can be found in~\cite{bd},
but are repeated here to set notation and for extension to string theory.

\subsection{The Setup and Vertex Operators} \label{sec:rindlerIntro}

Consider the $d$-dimensional Rindler spacetime with metric
\begin{equation} \label{rc}
\begin{gathered}
ds^2 = e^{2 a \xi} (-d\eta^2 + d\xi^2) + d\vec{y}^2.
\begin{aligned}
\end{aligned}
\end{gathered}
\end{equation}
Here $a$ is a constant, 
$\eta$ is time, $\xi$ is a spatial coordinate and $\vec{y}$ are $d-2$
additional, flat, spatial coordinates.  There is a horizon at
$\xi=-\infty$; $\xi=+\infty$ is spatial infinity.
Indeed, the
geometry becomes somewhat more illuminating via the coordinate
transformation
\begin{equation} \label{mc}
\begin{aligned}
t &= a^{-1} e^{a \xi} \sinh (a \eta), & \qquad
x &= a^{-1} e^{a \xi} \cosh (a \eta),
\end{aligned}
\end{equation}
which brings the metric to the form
\begin{equation}
ds^2 = -dt^2 + dx^2 + d\vec{y}^2,
\end{equation}
\ie\ Minkowski space,
for Rindler space is the spacetime seen by
an accelerating observer in a Minkowski background.

\FIGURE[t]{
\includegraphics[height=2in,clip=true,
      keepaspectratio=true]{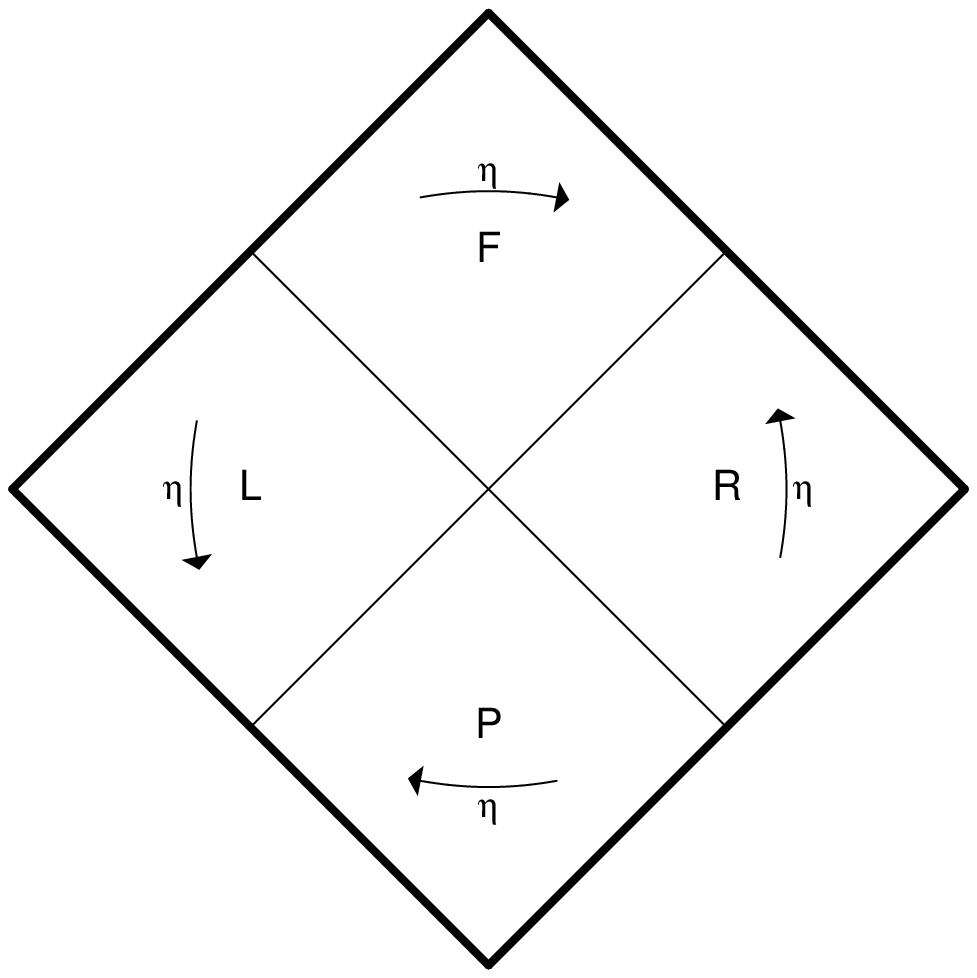}
\caption{Rindler space.  R is the Rindler region, L is the opposite
Rindler region, F is the future and P is the past.
The direction of time ($\eta$) in each region is also shown.
The diagonal lines are
the past and future Rindler horizons.  Each point depicts an
$\ZR^{d-2}$.\label{fig:rindler}}
}

Rindler space is plotted in Fig.~\ref{fig:rindler}.  The region R is
that covered by the coordinates~\eqref{rc}.  The region L is the other
Rindler region corresponding to $t\rightarrow -t$ and $x\rightarrow
-x$ in~\eqref{mc}, or equivalently, $\im \eta = -\pi i a^{-1}$.  Other
regions not covered by the coordinates~\eqref{rc} are the future (F)
and past (P).  These correspond to $\im \xi=\frac{\pi}{2a }$ with 
respectively $\im \eta = -\frac{\pi}{2a}$, and
$\im \eta = -\frac{3 \pi}{2 a}$.  
Equivalently, the other regions are covered via
\begin{gather}
\label{mcF}
\begin{aligned}
t &= a^{-1} e^{a \xi_F} \cosh (a \eta_F), & \qquad
x &= a^{-1} e^{a \xi_F} \sinh (a \eta_F),
\end{aligned} \\
\label{mcP}
\begin{aligned}
t &= -a^{-1} e^{a \xi_P} \cosh (a \eta_P), & \qquad
x &= -a^{-1} e^{a \xi_P} \sinh (a \eta_P),
\end{aligned} \\
\label{mcL}
\begin{aligned}
t &= -a^{-1} e^{a \xi_L} \sinh (a \eta_L), & \qquad
x &= -a^{-1} e^{a \xi_L} \cosh (a \eta_L),
\end{aligned}
\end{gather}
which can be summarized via
\begin{align}
\eta &= \eta_L - \frac{\pi}{a} i, \xi=\xi_L, &
\eta &= \eta_F - \frac{\pi}{2 a} i, \xi = \xi_F+\frac{\pi}{2a}i, &
\eta &= \eta_P - \frac{3\pi}{2 a} i, \xi = \xi_P+\frac{\pi}{2a}i,
\end{align}
Note the periodicity
of~\eqref{mc} under $\eta\rightarrow \eta - 2 \pi i a^{-1}$.  Also,
note that ``time translation'', defined with respect to $\eta$, goes
backwards in region L and is spacelike in the future and past.  This
is analogous to a black hole of temperature $a$.

Now consider a mass $\mu$
scalar field on Rindler space.  The wave equation reads
\begin{equation}
\left[e^{-2 a \xi} (-\p_\eta^2+\p_\xi^2) + \vec{p}_y^2 - \mu^2 \right] \phi = 0.
\end{equation}
The positive frequency modes are
\begin{equation}
\phi^R_{\omega,\vk} = \sqrt{\frac{\sinh\left(\frac{\pi \omega}{a}\right)}{a}}
   \frac{2}{(2\pi)^{d/2}} 
   e^{-i \omega \eta + i \vk \cdot \vec{y} }
   K_{i \frac{\omega}{a}} 
       \left( \frac{\sqrt{\mu^2+\vk^2}}{a} e^{a \xi} \right),
\end{equation}
where $K_\nu(z)$ is a modified Bessel function.  (The second solution
blows up at infinity, and so is excluded.)
These are (Klein-Gordon)
$\delta$-function normalized,
\begin{equation}
i \int_{-\infty}^\infty d\xi \int d^{d-2} \vec{y} \left[
   \phi^{R^*}_{\omega,\vk} \frac{\overleftrightarrow{\p}}{\p \eta}
   \phi^R_{\omega',\vk'} \right]
= \delta(\omega-\omega') \delta^{(d-2)}(\vk-\vk').
\end{equation}
These Rindler modes can be analytically continued into the future and
past (Fig.~\ref{fig:rindler}) and are defined to vanish in the other
(left) Rindler region.  One similarly defines positive frequency modes
in the left Rindler region.

As pointed out in $\S$\ref{sec:pinch}, the (tachyon) vertex operators of
the (bosonic) string are essentially the on-shell field theory
modes.  In Minkowski space, the vertex operators are (suppressing
the normal ordering symbol) $e^{ik X}=
e^{-i E t + i \ell x + i \vec{k} \cdot \vec{y}}$, with $k^2 = -\mu^2 =
\frac{1}{\alpha'}$ for the open string.
For the right Rindler region, we can use the formula%
\footnote{In addition to suppressing the normal ordering symbols, we also
suppress the worldsheet
dependence of the vertex operator which
appears via the string coordinates $t$, $x$ and $\vec{y}$.}
\begin{equation} \label{gotMink}
\begin{split}
{\mathcal V}^R_{\omega,\vec{k}}
&= \sqrt{\frac{\sinh\left(\frac{\pi \omega}{a}\right)}{a}}
   \frac{2}{(2\pi)^{d/2}}
 e^{-i \omega \eta + i \vec{k}\cdot \vec{y}}
 K_{i \frac{\omega}{a}}\left(\tfrac{\sqrt{\mu^2+\vk^2}}{a}\,e^{a \xi}\right)
\\ &= \frac{1}{2 (2\pi)^{d/2} \sqrt{ a\sinh \frac{\pi \omega}{a}}} 
\int_{-\infty}^\infty \frac{d\ell}{\sqrt{\mu^2+\vk^2+\ell^2}}
\left\{ e^{\frac{\pi \omega}{2a} - i \frac{\omega}{a}
    \sinh^{-1} \frac{\ell}{\sqrt{\mu^2 + \vk^2}}
  - i \sqrt{\mu^2+\ell^2+\vk^2} t + i \ell x + i \vec{k}\cdot \vec{y}}
\right. \\ & \hspace{.3\textwidth} \left.
- \; e^{-\frac{\pi \omega}{2a} +  i \frac{\omega}{a}
    \sinh^{-1} \frac{\ell}{\sqrt{\mu^2 + \vk^2}} 
  + i \sqrt{\mu^2+\ell^2+\vk^2} t + i \ell x + i \vec{k}\cdot \vec{y}}
\right\},
\end{split}
\end{equation}
to write the on-shell Rindler modes in terms of on-shell Minkowski ones.
Thus,
the right-hand side of~\eqref{gotMink} is a well-defined expression
for the tachyon vertex operator in the right Rindler region.

The left Rindler region vertex operators are similarly
\begin{multline} \label{gotLMink}
{\mathcal V}^L_{\omega,\vec{k}}
= \frac{1}{2 (2\pi)^{d/2}\sqrt{a \sinh \frac{\pi \omega}{a}}}
\int_{-\infty}^\infty \frac{d\ell}{\sqrt{\mu^2+\vk^2+\ell^2}}
\left\{ e^{\frac{\pi \omega}{2a} + i \frac{\omega}{a}
    \sinh^{-1} \frac{\ell}{\sqrt{\mu^2 + \vk^2}}
  - i \sqrt{\mu^2+\ell^2+\vk^2} t + i \ell x + i \vec{k}\cdot \vec{y}}
\right. \\ \left.
- \; e^{-\frac{\pi \omega}{2a} -  i \frac{\omega}{a}
    \sinh^{-1} \frac{\ell}{\sqrt{\mu^2 + \vk^2}} 
  + i \sqrt{\mu^2+\ell^2+\vk^2} t + i \ell x + i \vec{k}\cdot \vec{y}}
\right\}.
\end{multline}

Having written the vertex operators in terms of Minkowski ones, it is now
a relatively simple matter to compute correlation functions from the
Minkowski space ones.

\subsection{Rindler Space $\alpha$-Vacua} \label{sec:rindlerAlpha}

It is convenient to define the linear combinations of left and right Rindler
region modes
\begin{equation} \label{mRm}
\begin{aligned}
\phi^1_{\omega,\vk} &= \frac{1}{\sqrt{2}} \left(
   \phi^R_{\omega,\vk} + i \phi^L_{\omega,\vk}
   \right), \\
\phi^2_{\omega,\vk} &= \frac{1}{\sqrt{2}} \left(
   \phi^L_{\omega,\vk} + i \phi^R_{\omega,\vk}
   \right).
\end{aligned}
\end{equation}
These modes are orthonormal---and, in particular, $\phi^1$ are
orthogonal to $\phi^2$---if the Rindler modes are.
As these combinations only involve positive frequencies, the vacuum
defined by the modes~\eqref{mRm} is again the Rindler vacuum.

Now consider the modes
\begin{equation}
\begin{aligned}
\phi^{1(\alpha,\beta)}_{\omega,\vk}
&= \cosh \alpha \, \phi^1_{\omega,\vk}
    + e^{i \beta} \sinh \alpha \, 
            \phi^{1*}_{\omega,-\vk}, \\
\phi^{2(\alpha,\beta)}_{\omega,\vk}
&= \cosh \alpha \, \phi^2_{\omega,\vk}
    + e^{i \beta} \sinh \alpha \,
             \phi^{2*}_{\omega,-\vk},
\end{aligned}
\end{equation}
Again, these modes are orthonormal.
These modes define the $\alpha$-vacuum
$\ket{\alpha,\beta}$.

It must be emphasized that unlike the black hole $\alpha$-vacua,
the identification (\cf.~$\S$\ref{sec:inv}) associated with this definition
of Rindler space $\alpha$-vacua is afflicated with a fixed point at
the origin of Minkowski space.

\subsection{No Pinch Singularities in One-loop Diagrams}
\label{sec:rindlerLoop}

Let us now consider one-loop diagrams.
For simplicity, we consider bosonic open string diagrams with external tachyon
legs, and recall~\cite{jp} that the mass-squared of the tachyon is
$\mu^2=-\frac{1}{\alpha'}$.
The field theory diagram,
eq.~\eqref{oneloop}, suffered from a pinch singularity.  The open
string theory analog of this diagram is depicted
in~Fig.~\ref{fig:open2} (page~\pageref{fig:open2}).
It is expected to be nonsingular; in
particular, it is equivalent to a tree-diagram in which a single
closed string is exchanged.  However, the closed string picture is good for
the regime of modular parameter which is opposite to the field theory limit.
Thus, the mechanism in which the pinch singularity is resolved is inherently
stringy.

The simple analysis
performed here will be complicated by the fact that string theory diagrams 
are necessarily on-shell.  Thus, for example, the two-point function which
epitomized the pinch singularity in field theory is not easy to analyze
in string theory---the propagator is by its very nature singular on-shell.
So we will examine the three-point function at one-loop, 
Fig.~\ref{fig:3pt},
and see that
it also does not appear to have problems that signal a pinch singularity.
Rather, we will see how the pinch singularity is regulated.

\FIGURE[t]{
\begin{tabular}{cc}
\subfig{\includegraphics[width=3in,height=1in,keepaspectratio=true]{%
        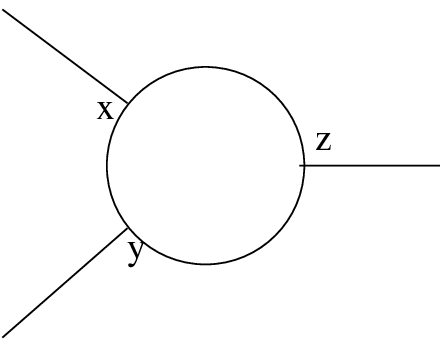}
        \label{subfig:3ptft}} &
\subfig{\includegraphics[width=3in,height=1in,keepaspectratio=true]{%
        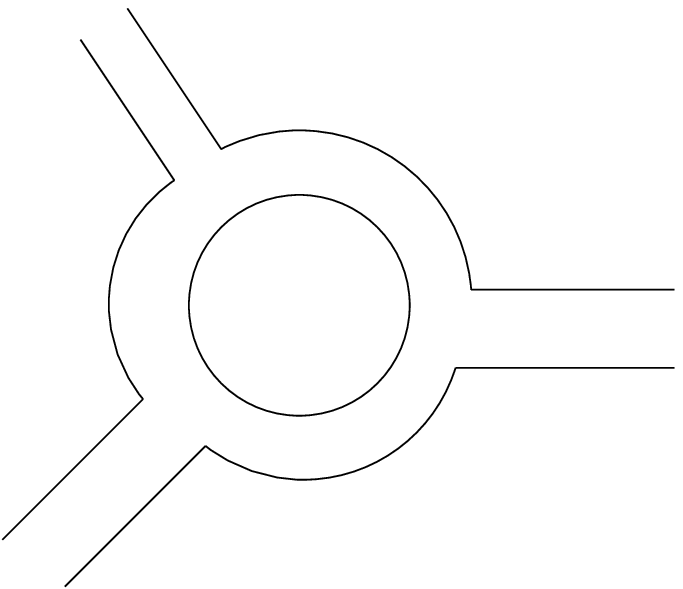}
        \label{subfig:3ptst}}
\end{tabular}
\caption{The 1-loop contribution to the three point function in
\ref{subfig:3ptft} field theory and \ref{subfig:3ptst} the planar
contribution to string theory.  As position-space diagrams,
the pinch singularities appear for $\alpha$-vacua in field theory,
when, say, both the integrated points $x$ and $y$, and $y$ and $z$ are 
null-separated, so that the $y$-integration has a double pole, with potentially
conflicting $i\epsilon$ prescriptions.  The string theory diagram shows
that this is smoothed out.\label{fig:3pt}}
}

\subsubsection{Minkowski space diagrams} \label{sec:rindlerMinkLoop}

We will need Minkowski space one-loop amplitudes.
The general result is well-known~\cite{jp};
we have adapted it from~\cite{lm}.
In terms of the open-string modular parameter,
$t$, the ratio between the radius and the length of the
cylinder, and the vertex operator positions $\tau$, one finds that
the $M$-point amplitude with the first $N$ vertex operators
on one boundary and the remaining $M-N$ on the other boundary, is
\begin{multline}
{\mathcal A}(k_1,\dots,k_N;k_{N+1},\dots,k_M)
\\ = i g^M \delta^{(26)}(\sum_{p=1}^M k_p) \int_0^\infty \frac{dt}{2t}
(8 \pi^2 \apr t)^{-13} \eta(it)^{-24} 
\left[\prod_{p=1}^M \int_0^{2\pi t} d\tau_p\right] \Psi_1 \Psi_2 \Psi_{12},
\end{multline}
where
\begin{equation}
\begin{gathered}
\begin{aligned}
\Psi_1 &= \prod_{1\leq i<j\leq N} \abs{\psi_{ij}}^{2\apr k_i\cdot k_j},
&
\Psi_2 &= \prod_{N+1\leq r<s\leq M} \abs{\psi_{rs}}^{2\apr k_r\cdot k_s},
&
\Psi_{12} &= \prod_{\substack{1\leq i\leq N \\ N+1\leq r \leq M}} 
   \left(\psi^T_{ir}\right)^{2\apr k_i\cdot k_r},
\end{aligned} \\
\begin{aligned}
\psi_{ij} &= -2\pi i \exp \left(-\frac{\tau_{ij}^2}{4\pi t}\right)
   \frac{\vartheta_{11}\left(i \frac{\tau_{ij}}{2\pi}|it\right)}{
         \vartheta'_{11}(0|it)},
&
\psi^T_{ir} &= 2\pi \exp \left(-\frac{\tau_{ir}^2}{4\pi t}\right)
   \frac{\vartheta_{10}\left(i \frac{\tau_{ir}}{2\pi}|it\right)}{
         \vartheta'_{11}(0|it)},
\end{aligned}
\end{gathered}
\end{equation}
$\tau_{ij}=\tau_i-\tau_j$, $\vartheta$ and $\eta$ are Jacobi
$\vartheta$- and Dedekind $\eta$-functions using the conventions of~\cite{jp},
and $g$ is the open string coupling.

\subsubsection{The Three-Point Function} \label{sec:rindler3}

A quantity
which we expect to contribute to a pinch singularity is
the one-loop contribution to
$\vev{\phi^R_{\omega_1,\vec{k}_1} \phi^R_{\omega_2,\vec{k}_2}
      \phi^L_{\omega_3,\vec{k}_3}}$.
This is given, up to factors of $2$, $\pi$, and $g$, by
\begin{subequations}
\begin{multline}
{\mathcal A}_{RRL}
= \frac{1}{\sqrt{\sinh\frac{\pi\omega_1}{a} \sinh\frac{\pi\omega_2}{a}
           \sinh\frac{\pi\omega_3}{a}}}
  \int_{-\infty}^\infty \frac{d\ell_1}{\sqrt{\mu_1^2+\ell_1^2}}
  \int_{-\infty}^\infty \frac{d\ell_2}{\sqrt{\mu_2^2+\ell_2^2}}
  \int_{-\infty}^\infty \frac{d\ell_3}{\sqrt{\mu_3^2+\ell_3^2}}
\\ \times
  \sum_{\sigma_1,\sigma_2,\sigma_3=\pm} \sigma_1 \sigma_2 \sigma_3
  e^{\frac{\pi}{2a} (\sigma_1\omega_1+\sigma_2\omega_2+\sigma_3\omega_3)}
  e^{-i\sigma_1 \frac{\omega_1}{a} \sinh^{-1} \frac{\ell_1}{\mu_1}
     -i\sigma_2 \frac{\omega_2}{a} \sinh^{-1} \frac{\ell_2}{\mu_2}
     +i\sigma_3 \frac{\omega_3}{a} \sinh^{-1} \frac{\ell_3}{\mu_3}}
\\ \times
  \delta\left(
         \sigma_1 \sqrt{\mu_1^2+\ell_1^2} + \sigma_2 \sqrt{\mu_2^2+\ell_2^2}
         + \sigma_3 \sqrt{\mu_3^2+\ell_3^2}\right)
  \delta(\ell_1+\ell_2+\ell_3) \delta(\vec{k}_1+\vec{k}_2+\vec{k}_3)
\\ \times
  \int_0^\infty \frac{dt}{2t} (8\pi\alpha' t)^{-13} \eta(it)^{-24}
  \int_0^{2\pi t} d\tau_1 \int_0^{2\pi t} d\tau_2 \int_0^{2\pi t} d\tau_3
\,  e^{-\frac{\alpha'}{2\pi t} \sum_{i<j} k_i\cdot k_j \tau_{ij}^2}
\\ \times
  \abs{2\pi \frac{\vartheta_{1\cdot}(i\frac{\tau_{12}}{2\pi}|it)}{
             \vartheta_{11}'(0|it)}}^{2\alpha' k_1\cdot k_2}
  \abs{2\pi \frac{\vartheta_{1\cdot}(i\frac{\tau_{13}}{2\pi}|it)}{
             \vartheta_{11}'(0|it)}}^{2\alpha' k_1\cdot k_3}
  \abs{2\pi \frac{\vartheta_{1\cdot}(i\frac{\tau_{23}}{2\pi}|it)}{
             \vartheta_{11}'(0|it)}}^{2\alpha' k_2\cdot k_3}.
\end{multline}
where the $\sigma_i$ are associated with the signs in the exponentials 
in the two terms of~\eqref{gotMink} and~\eqref{gotLMink}; the missing
index in the $\vartheta$-functions is $1$ or $0$ depending on whether
the associated vertex operators are on the same or different boundaries;
it is understood that
$k_i = (\sigma_i \sqrt{\mu_i^2+\ell_i^2}, \ell_i, \vec{k}_i)$;
and we have made the convenient definition
\begin{equation} \label{defmui}
\mu_i = \sqrt{\mu^2 + \vec{k}_i^2}, \qquad \mu^2 = -\frac{1}{\alpha'},
\end{equation}
\end{subequations}
in terms of the spatial momentum $\vec{k}_i$ at each vertex, and the
tachyon mass.
In fact, the integrals are simplified by using the effective
kinematics associated with a 1+1-dimensional 3-particle scattering for
particles of mass $\mu_i$ and momentum $\ell_i$.

The energy-momentum conserving $\delta$-functions
imply that one of the $\sigma$'s must be
different from the other two.
Consider the term for which the $\sigma_1=\sigma_2=-\sigma_3$; 
we will obtain the other terms by permutation.
We then use the $\delta$-functions
to integrate over and eliminate $\ell_1$ and $\ell_2$.  The
$\ell_3$ integration is simplified by defining
$\zeta=\sinh^{-1} \frac{\ell_3}{\sigma_3 \mu_3}$, which is the boost parameter
required to transform the kinematics from the $\ell_3=0$ frame of
the effective 1+1-dimensional scattering problem.
One then finds that the $\delta$-functions impose
\begin{subequations}
\begin{align} \label{gotell}
\ell_1 &= \mu_1 \sinh(\sigma_1 \zeta \pm \lambda_1), &
\ell_2 &= \mu_2 \sinh(\sigma_2 \zeta \mp \lambda_2),
\end{align}
where the arbitrary sign
is the $1+1$-dimensional analogue of the arbitrary
angle which appears in higher-dimensional scattering, and
\begin{align} \label{lambdas}
\lambda_1 &\equiv \sinh^{-1} \frac{\kappa^2}{2 \mu_1 \mu_3}, &
\lambda_2 &\equiv \sinh^{-1} \frac{\kappa^2}{2 \mu_2 \mu_3}, &
\kappa^4 &\equiv \mu_1^4 + \mu_2^4 + \mu_3^4 
  - 2 \mu_1^2 \mu_2^2 - 2 \mu_1^2 \mu_3^2 - 2 \mu_2^2 \mu_3^2.
\end{align}
\end{subequations}
Therefore,
\begin{subequations} \label{kdotk}
\begin{align}
k_1\cdot k_2 &= -\mu_1\mu_2 \cosh(\lambda_1 + \lambda_2) 
 + \vec{k}_1\cdot \vec{k}_2, \\
k_1\cdot k_3 &= \mu_1 \mu_3 \cosh \lambda_1 + \vec{k}_1 \cdot \vec{k}_3, \\
k_2\cdot k_3 &= \mu_2 \mu_3 \cosh \lambda_2 + \vec{k}_2 \cdot \vec{k}_3,
\end{align}
\end{subequations}
which is independent of both $\zeta$ and the sign in~\eqref{gotell}.
The $\ell_3$---or rather $\zeta$---integration then yields the 
expected $\omega$-constraining $\delta$-function; the final result, upon
summing over the arbitrary sign in~\eqref{gotell}, is thus
\begin{multline} \label{simpA3}
{\mathcal A}_{RRL}
= \frac{1}{
        \sqrt{\sinh\frac{\pi\omega_1}{a} \sinh\frac{\pi\omega_2}{a}
         \sinh\frac{\pi\omega_3}{a}}}
\delta(\omega_1+\omega_2-\omega_3) \delta(\vec{k}_1+\vec{k}_2+\vec{k}_3)
\\ \times
\left \{
2 \frac{\sinh\left[\frac{\pi}{2a}(\omega_1+\omega_2-\omega_3)\right]
        \cos (\lambda_1-\lambda_2)}{\mu_1\mu_2 \sinh(\lambda_1+\lambda_2)}
  \int_0^\infty \frac{dt}{2t} (8\pi\alpha' t)^{-13} \eta(it)^{-24}
\right. \\ \times \left.
  \int_0^{2\pi t} d\tau_1 \int_0^{2\pi t} d\tau_2 \int_0^{2\pi t} d\tau_3
  e^{-\frac{\alpha'}{2\pi t} 
     k_1\cdot k_2
     \tau_{12}^2}
  \abs{2\pi\frac{\vartheta_{1\cdot}(i\frac{\tau_{12}}{2\pi}|it)}{
             \vartheta_{11}'(0|it)}}^{2\alpha' 
          k_1\cdot k_2
           }
\right. \\ \times \left.
  e^{-\frac{\alpha'}{2\pi t} 
        k_1\cdot k_3
     \tau_{13}^2}
  \abs{2\pi\frac{\vartheta_{1\cdot}(i\frac{\tau_{13}}{2\pi}|it)}{
             \vartheta_{11}'(0|it)}}^{2\alpha' 
          k_1 \cdot k_3
          }
  e^{-\frac{\alpha'}{2\pi t} 
          k_2\cdot k_3
          \tau_{13}^2}
  \abs{2\pi\frac{\vartheta_{1\cdot}(i\frac{\tau_{23}}{2\pi}|it)}{
             \vartheta_{11}'(0|it)}}^{2\alpha' 
        k_2 \cdot k_3
        }
\right. \\ \left.
+ \text{2 perms}
\right\},
\end{multline}
where $k_i\cdot k_j$ are given in terms of the $\vec{k}$'s and
the masses by~\eqref{kdotk}.
The sign in the energy conserving $\delta$-function is due to the leftness
of the Rindler mode, not the relative value of $\sigma_3$, but the
sign in the $\sinh$ in the numerator is due to the relative value of
$\sigma_3$.
Because the $\lambda$'s and $\mu$'s depend only on the $\vec{k}$'s, 
and appear in the amplitude~\eqref{simpA3} via $k_i\cdot k_j$ for on-shell
$k$'s, the final integrations to be performed are identical to those for
Minkowski space amplitudes, which are well-known to be well-behaved.  Thus
we see that the stringy 
amplitudes for $\alpha$-vacua in Rindler space are also
well-behaved.

\paragraph{Open String Factorization} \label{sec:osfac}
It is the large $t$ regime of~\eqref{simpA3} from which 
the field theory pinch singularities are expected.
Approximating the integrand of the planar diagram
at large $t$, and, using the translational symmetry of
the annulus to fix $\tau_1$ to a convenient value yields, (\cf\ \eg~\cite{lm})
\begin{subequations}
\begin{multline} \label{A3osfac1}
{\mathcal A}_{RRL}
= \frac{1}{
        \sqrt{\sinh\frac{\pi\omega_1}{a} \sinh\frac{\pi\omega_2}{a}
         \sinh\frac{\pi\omega_3}{a}}}
\delta(\omega_1+\omega_2-\omega_3) \delta(\vec{k}_1+\vec{k}_2+\vec{k}_3)
(1-y_1)^2
\\ \times
\left \{
2 \pi \frac{\sinh\left[\frac{\pi}{2a}(\omega_1+\omega_2-\omega_3)\right]
            \cos (\lambda_1-\lambda_2)}{\mu_1\mu_2 \sinh(\lambda_1+\lambda_2)}
  \int_0^\infty dt (8\pi\alpha' t)^{-13} e^{-2\pi t}
\right. \\ \times \left.
  \int_{-y_0}^{y_0} dy_2 \int_{-y_0}^{y_0} dy_3\,
  e^{-\frac{\alpha'}{2\pi t} \sum_{i<j} k_i\cdot k_j
     \tau_{ij}^2}
  \abs{y_{12}}^{2\alpha' 
          k_1\cdot k_2
           }
  \abs{y_{13}}^{2\alpha' 
          k_1 \cdot k_3
          }
  \abs{y_{23}}^{2\alpha' 
        k_2 \cdot k_3
        }
+ \text{2 perms}
\right\},
\end{multline}
where 
\begin{align}
\tau_i &= -\ln \abs{\frac{1+y_i}{1-y_i}} + \pi t, &
y_0 &= \tanh \frac{\pi t}{2},
\end{align}
\end{subequations}
The difference between the planar and the nonplanar diagram,
in this approximation, is that the $y$'s on the ``other'' boundary
would lie in \hbox{$(-\infty,-y_0^{-1})\cup(y_0^{-1},\infty)$} instead
of $(-y_0,y_0)$.  In fact, in the $t\rightarrow\infty$ approximation,
$y_0\rightarrow 1$.  Incorporating this,
and introducing a new integration for reasons to become clear,
the amplitude~\eqref{A3osfac1} can be rewritten
as
\begin{multline} \label{A3osfac2}
{\mathcal A}_{RRL}
= \frac{1}{
        \sqrt{\sinh\frac{\pi\omega_1}{a} \sinh\frac{\pi\omega_2}{a}
         \sinh\frac{\pi\omega_3}{a}}}
\delta(\omega_1+\omega_2-\omega_3) \delta(\vec{k}_1+\vec{k}_2+\vec{k}_3)
(1-y_1^2)
\\ \times
\left \{
2 \frac{\sinh\left[\frac{\pi}{2a}(\omega_1+\omega_2-\omega_3)\right]
        \cos (\lambda_1-\lambda_2)}{\mu_1\mu_2 \sinh(\lambda_1+\lambda_2)}
  \int_0^1 \frac{dq}{2 q} \int \frac{d^{26}k}{(2\pi)^{26}}
  q^{\alpha'(k^2+\mu^2)}
  \int_{-1}^{1} dy_2 \int_{-1}^{1} dy_3
\right. \\ \times \left.
 e^{-2\alpha' k \cdot (k_1 \tau_1 + k_2 \tau_2 + k_3 \tau_3)}
  \abs{y_{12}}^{2\alpha' 
          k_1\cdot k_2
           }
  \abs{y_{13}}^{2\alpha' 
          k_1 \cdot k_3
          }
  \abs{y_{23}}^{2\alpha' 
        k_2 \cdot k_3
        }
+ \text{2 perms}
\right\},
\end{multline}
where $q=e^{-2\pi t}$.
Integration of the Gaussian integral $k$ yields the original
expression~\eqref{A3osfac1}.  However, the form~\eqref{A3osfac2},
shows the open string factorization illustrated in Fig.~\ref{fig:openfac}
(page~\pageref{fig:openfac}).
In particular, $y_i$'s are the vertex operator positions on the
upper half-plane and $\pm k$ are the momenta of the extra vertex operator
insertions.  (The fixing of $y_1=0$, as well as the fixing of the extra
vertex operator positions at $y=\pm 1$ is maximal fixings
allowed by the conformal Killing vectors of the upper half-plane.)

If one integrates $q$ instead of $k$ in~\eqref{A3osfac2},
a factor of $\frac{1}{k^2+\mu^2}$, the field theory propagator of the
intermediate state, is obtained.  Moreover, there is a pole that involves
$k+k_3$ via the factors of 
$e^{-2\alpha' k \cdot k_3 (\tau_3 - \pi t)} = 
\abs{\frac{1+y_3}{1-y_3}}^{-\alpha' (k+ k_3)^2 +\alpha' k^2 - 1}$
(where all $\tau_i$'s could be replaced by $\tau_i - \pi t$ due to
energy-momentum conservation, and using $\alpha'\mu^2=-1$).
Thus, exchange of a tachyon ($k^2=-\mu^2=\frac{1}{\alpha'}$)
leads to simultaneous poles which, because the second pole is associated
with $y_3$---\ie\ the term from the left Rindler region which therefore
has the opposite $i\epsilon$ prescription---gives
the pinch singularity.  However, we have already seen from
the complete result~\eqref{simpA3} that these 
apparent pinch singularities
smoothed out by the full string theory.  This is because the
$y_3$ integration only goes to $\pm y_0$, not $\pm 1$ where the singularity
appears.

\end{document}